\documentclass[12pt]{iopart}   
\usepackage{iopams}
\usepackage{graphicx}
\usepackage{amssymb}
 \usepackage{color}
\usepackage[english]{babel}

\begin{document}

\title[Thermal distortions of non-Gaussian beams]{Thermal distortions
  of non-Gaussian beams in Fabry-Perot cavities}

\author{J Miller$^{1,2}$, P Willems$^2$, H Yamamoto$^{2}$, J Agresti$^{2,3}$\\ and R DeSalvo$^2$}
\address{$^1$IGR, University of
  Glasgow, University Avenue, Glasgow, G12 8QQ, U.K.}
\address{$^2$LIGO Laboratory, California Institute of Technology, 1200
  E. California Blvd., Pasadena, CA 91125, USA}
\address{$^3$Dipartimento di Fisica presso Universit\`a di Pisa, via
  Fermi 8, 56126 Pisa, Italy}
\ead{j.miller@physics.gla.ac.uk}
\submitto{\CQG}
\maketitle

\begin{abstract}Thermal effects are already important in currently
  operating interferometric gravitational wave detectors.  Planned
  upgrades of these detectors involve increasing optical power to
  combat quantum shot noise. We consider the ramifications of this
  increased power for one particular class of laser beams - wide,
  flat-topped, mesa beams. In particular we model a single mesa beam
  Fabry-Perot cavity having thermoelastically deformed mirrors. We
  calculate the intensity profile of the fundamental cavity eigenmode
  in the presence of thermal
  perturbations, and the associated changes in thermal noise.
  We also outline an idealised method of correcting
  for such effects. At each stage we contrast our results with those
  of a comparable Gaussian beam cavity. Although we focus on mesa
  beams the techniques described are applicable to any azimuthally
  symmetric system.
\end{abstract}

\section{Introduction}

The sensitivity of current kilometre-scale interferometric
gravitational wave detectors such as GEO600 \cite{GEO600}, Virgo
\cite{Virgo}, and LIGO \cite{LIGO} is limited by fundamental noise
processes.  One of these is shot noise in the detected light power and
for this reason they operate with kilowatts of stored power. Planned
improvements to these detectors will increase this stored power to the
hundreds of kilowatts range.

Another fundamental limit to sensitivity is thermal motion of the
interferometer mirrors.  It is anticipated that the dominating noise
source in the middle of the terrestrial gravitational wave detection
band will be coating thermal noise \cite{Harry06}. These
thermodynamical effects cause the surface of a test mass to
fluctuate stochastically on a microscopic scale. Crudely speaking,
interferometric gravitational wave detectors operate by measuring
the position of their test masses' high reflectivity surfaces
weighted by the intensity profile of the arm cavity eigenmode.
Narrow, sharply peaked Gaussian beams which meet 1 ppm diffraction
loss requirements are not optimal - they provide a poor spatial
average of these thermal fluctuations. Heuristically, a wider beam
with a more uniform intensity profile will average over a larger
number of fluctuations and thus reduce the impact of test mass
thermal noise. One such beam which has been proposed for use in
gravitational wave interferometers is the mesa beam. This beam has
been predicted to reduce mirror thermal noise by around a factor of
two relative to a Gaussian beam, without being significantly more
difficult to control
\cite{D'Ambrosio03,D'Ambrosio04,O'Shaughnessy04,Savov06}.  The mesa
beam resonates in Fabry-Perot cavities with specially tailored
aspherical mirror surfaces.

The LIGO detector already employs a thermal compensation system to
correct some mirrors' radial profiles against thermal effects arising
from absorption of stored optical power \cite{Ballmer05}. In future
high power upgrades, thermal perturbations will be commensurately
increased. They will distort the mirror surfaces, changing the
structure of the resonant optical mode. In turn this will change the
measured thermal noise, and potentially reduce the stored power due to
scattering of light out of the fundamental mode of the arm cavities or
by degrading the coupling with the injected beam.

We study thermally induced perturbations of a Fabry-Perot cavity in
the presence of high circulating power and consider how a
thermoelastically distorted test mass affects the intensity profile of
the resonant optical mode. We evaluate the thermal noise performance
of the new eigenmode and discuss possible methods of compensating for
the deformed test masses.

Two cavities supporting non-Gaussian mesa beams are considered, one
nearly flat \cite{D'Ambrosio03,Tarallo07}, the other nearly concentric
\cite{Bondarescu06}. As a foil to these cases we also study a nearly
concentric spherical cavity. All three cavities have a length of 4
km. Each mirror of the spherical cavity has a radius of curvature 2076
m and therefore supports well-known Gaussian modes. These parameters
are similar to the proposed Advanced LIGO baseline configuration. For
each cavity the input beam is that Gaussian beam which is optimally
coupled to the unperturbed or `cold' cavity. This injection beam
remains fixed for all calculations in each case.

\subsection{Intensity/Mirror profiles}

In the unperturbed spherical mirror cavity, the resonant optical field
at the mirror surface is a fundamental Gaussian mode. The phase fronts
of this beam, and therefore the cavity mirrors, are spherical. For a
discussion of Gaussian beams and their properties see, for example,
Kogelnik \& Li \cite{Kogelnik66}.

In the nearly flat cavity case the unnormalised mesa field at the
mirror position is given by \cite{Tarallo07},
\begin{eqnarray}
\Psi_{\mathrm{mesa}}(r)&\propto\int_{\vec{r}\,'\leq D}\exp\bigg[-\frac{(\vec{r}-\vec{r}\,')^2(1-\rmi)}{2\omega_0^2}\bigg]\mathrm{d}^2\vec{r'}\nonumber\\
&=2\pi\int^{D}_{0}\exp\bigg[-\frac{(r^2+r'^2)(1-\rmi)}{2\omega_0^2}\bigg]I_0\bigg(\frac{rr'(1-\rmi)}{\omega_0^2}\bigg)r'\mathrm{d}r'
\end{eqnarray}
where $I_0(x)$ is a zeroth order modified Bessel function of the first
kind and $\omega_0=\sqrt{L/k}$ is the waist of the minimally spreading
Gaussian for that cavity ($L$ being the cavity length, $k$ the
wavenumber $2\pi/\lambda$, with $\lambda=$1064 nm).  We take the
radius of the disc over which we integrate to be $D=3.55\omega_0$.
This value gives a diffraction loss in the Advanced LIGO arm cavity of
approximately 0.5~ppm, as does the choice of spherical mirror
parameters in the Gaussian mode case above. Knowing the field we may
readily calculate the mirror profile $z_{\mathrm{HR}}$,
\begin{equation}
  z_{\mathrm{HR}}(r)=\frac{\mathrm{Arg}[\Psi_{\mathrm{mesa}}(r)]-\mathrm{Arg}[\Psi_{\mathrm{mesa}}(0)]}{k}
\end{equation}

Construction of the mirror profile in the nearly concentric mesa case
is expedited by the duality relations discovered by Agresti
\cite{Agresti05}. Using these relations one finds that the nearly
concentric mesa cavity mirror profile is nothing other than a
perfectly concentric sphere ($R=L/2$) with $z_{\mathrm{HR}}$
subtracted. At the mirror position this particular geometry gives rise
to the same intensity profile as the nearly flat cavity, more general
cavities are discussed by Bondarescu and Thorne \cite{Bondarescu06}.

\begin{figure}[htbp!]
\centering
\includegraphics[width=0.85\textwidth]{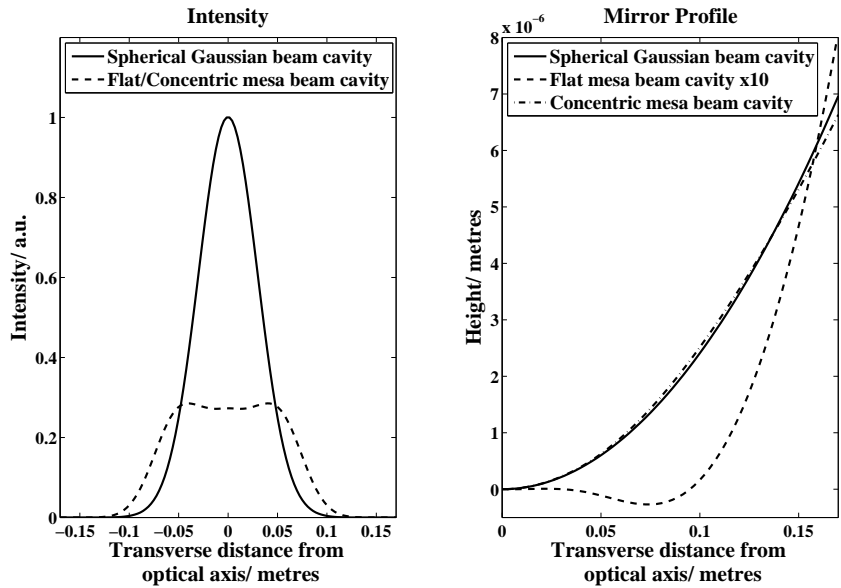}
\caption{A comparison of mesa (dashed lines) and Gaussian (solid
  lines) beams and the mirrors which support them. Left: Intensity
  profiles normalised to have equal power. The spot size of the
  Gaussian beam (where the intensity falls by $1/e^2$) is 6 cm, while
  that of the mesa beam is $\simeq 12$ cm; i.e. at FWHM the Gaussian
  beam samples $\sim$4\% of the mirror's surface whereas the mesa beam
  samples over 27\%. Right: Nominal mirror profiles for an AdvLIGO
  cavity. The flat mesa beam mirror profile has been expanded by a
  factor of ten to better show its structure. The concentric mesa
  mirror profile is realised by subtracting the flat mesa profile
  from a concentric sphere with $R=L_{\mathrm{cavity}}/2$. The abscissa
  extends to 0.17 m., the baseline mirror radius for
  AdvLIGO.} \label{fig:Comparison}
\end{figure}

Figure \ref{fig:Comparison} contrasts the intensity and mirror
profiles of Gaussian and mesa beams for an unperturbed Advanced LIGO
cavity.  The figure shows clearly the point made in the introduction
that the mesa beam samples more of the mirror surface ($\sim 50\%$)
than does the Gaussian beam of similar diffractive loss.

\section{Simulation tools}
\label{sec:SIS} Static Interferometer Simulation, SIS, is a program
developed at Caltech/LIGO in order to study, in detail, the optical
aspects of the Advanced LIGO interferometer \cite{YamamotoTech07}.  In
SIS, optical fields at mirror surfaces are evaluated over a spatial
grid. The fields are propagated from mirror to mirror by first
transforming them into a wave vector basis using a fast Fourier
transform and then propagating the transformed field from the first
mirror to the second using a paraxial approximation. At the second
mirror the optical field is transformed back into a spatial basis and
the transverse phase profile of the mirror applied to the
field. Optical fields combining from opposite sides of a mirror
surface are also summed at this point. The resolution of the
simulation is determined by the shortest spatial wavelength and can be
chosen as short as one needs so long as the paraxial approximation
holds. In the calculations done here, the grid was 256x256 pixels on a
0.7 m square. Checking our results against grid spacing, this
configuration was found to provide ample resolution for the smoothly
varying mirror profiles under study.

SIS uses an iterative procedure to find the stationary fields for a
given optical configuration and input field spectrum. The mirrors'
positions can be `locked' to the appropriate fraction of an optical
fringe by applying error signals calculated using standard heterodyne
techniques. SIS can also calculate the signal sideband induced by small
motions of the mirrors.  Surface deformations, such as thermal
deformation (using the Hello-Vinet approximation \cite{Vinet90}),
measured aberrations, randomly generated profile errors and
micro-roughness can also be included.

The choice of where to convert from Gaussian to mesa beams is not
obvious. In this article we assume that the arm cavity is driven by a
Gaussian input field. With 1070 W of power, this input gives 850 kW of
circulating power in the unperturbed spherical mirror cavity, a value
considered for Advanced LIGO. However, the spot size of the incident
Gaussian beam differs among the three configurations: for the
spherical mirror cavity it is 6 cm to match the mode resonating in the
unperturbed cavity. For the mesa beam cavities the input spot size is
8.4 cm in the flat case and 8.2 cm for the concentric system, to
optimise the coupling to the unperturbed cavity's mesa beam mode. This
optimised coupling is 95\% in both cases.  Thus the power build-up in
the unperturbed mesa cavities is only 808 kW\footnote{Thermorefractive
  aberrations will also be present in the input mirror substrate, but
  these have been ignored in this study so as to better understand the
  cavity effects.  This is equivalent to assuming that the purely
  thermorefractive aberrations have been compensated on the input
  field prior to injection into the cavity. It should be noted that
  such compensation is far from trivial.}.

\begin{figure}[htbp!]
\centering
\includegraphics[width=0.6\textwidth]{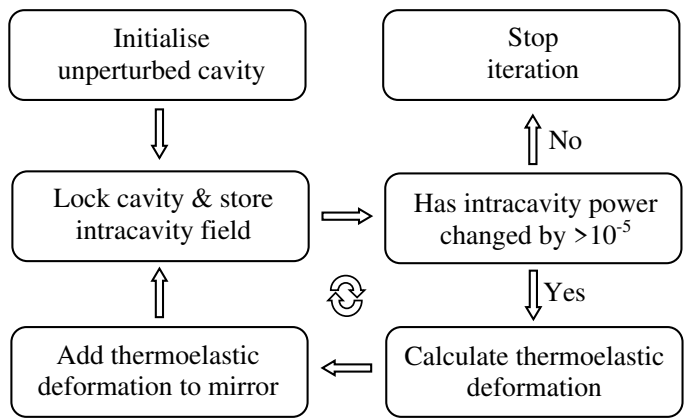}
\caption{Flowchart detailing the iteration procedure used to find
  static self-consistent intracavity fields and the thermoelastic
  deformations they produce.}
\label{fig:Flowchart}
\end{figure}

In this study we sought static, self-consistent solutions of the
optical fields and thermal deformations, using the following
procedure: starting with an unperturbed cavity and no stored optical
power, a field is injected through the input mirror, and the fields
throughout the cavity propagated and updated.  These fields are
re-propagated and updated iteratively until the stored intracavity
power is stable to one part in $10^5$ between successive
iterations. We then calculated the thermoelastic distortion of the
mirror surfaces for this intracavity mode shape and power and some
assumed mirror coating absorption\footnote{Absorption in the input
  mirror substrate is also generally present and also contributes to
  the thermoelastic deformation, but this contribution is negligible
  because the heating of the coating is much greater due to the high
  arm cavity finesse and the thermoelastic deformation of the mirror
  surface per unit absorbed power is much less for the substrate than
  coating \cite{LawrenceThesis}.}. This distortion was then added to
the mirror phase profile and the optical simulation restarted with no
stored intracavity power.  Again we seek a stable optical solution,
but generally with a different mode shape and stored power due to the
distortion.  The thermoelastic distortion due to this new intracavity
optical field was applied to the mirrors, and the procedure repeated
until the stored intracavity power is stable between the larger
distortion iterations to within one part in $10^5$. This process is
summarised in figure \ref{fig:Flowchart}.

\begin{figure}[htbp!]
\centering
\includegraphics[width=0.65\textwidth]{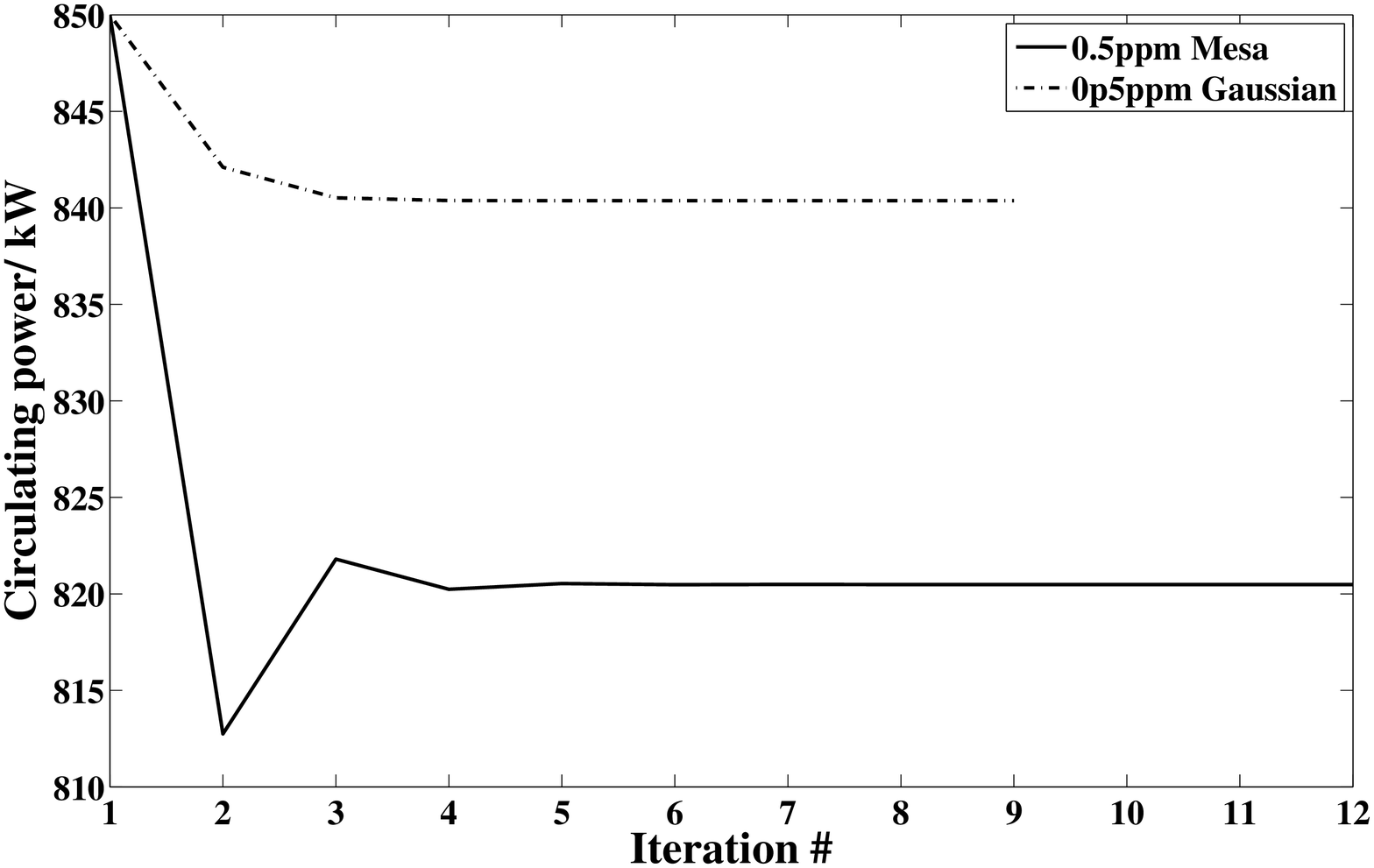}
\includegraphics[width=0.65\textwidth]{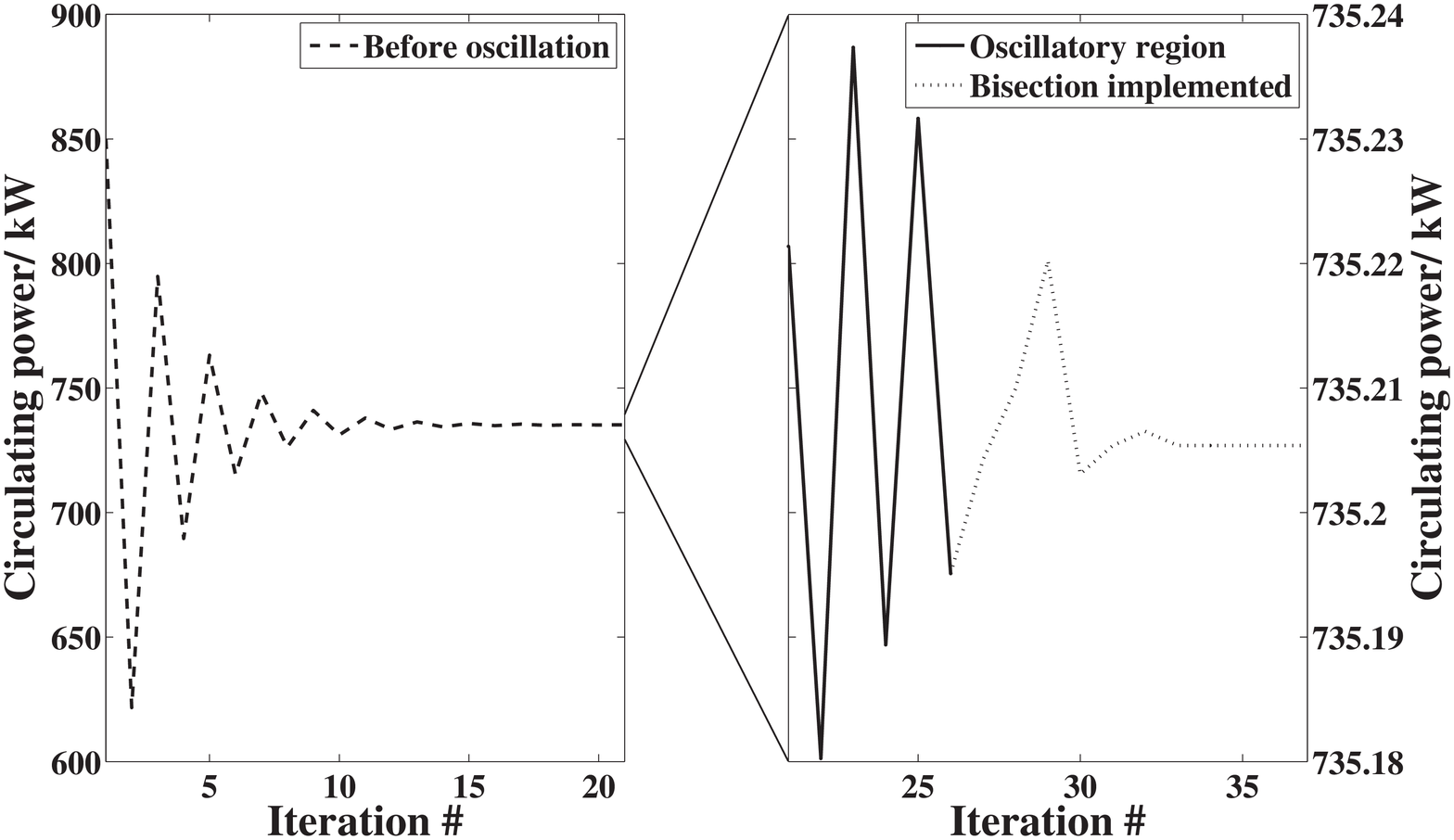}
\caption{Top: Intracavity power as a function of iteration number for
  0.5 ppm absorption. Convergence is achieved within 10 iterations in
  the Gaussian case; the mesa system requires 15 refinements. At
  higher absorption rates low level numerical oscillations were
  found. A simple bisection algorithm was implemented and rapid
  convergence achieved. The bottom panes demonstrate this in the flat
  mesa beam case for 2.5 ppm absorption. Bottom left: Gross
  convergence is achieved within 20 iterations. Bottom right: On
  closer inspection a low level numerical oscillation is
  present. Convergence is expedited by our bisection algorithm. To
  allow for easy comparison of different systems, the mesa intracavity
  powers have been normalised such that unperturbed mesa and Gaussian
  cavities store the same power. In reality the power stored in the
  mesa cavity is somewhat lower.}
\label{fig:Convergence}
\end{figure}

We found that for low powers convergence was achieved within 10
iterations; at higher powers convergence was slower. As the system
approached equilibrium the system was found to oscillate numerically
between distinct optical modes and thermal distortions.  These
problems tended to occur at larger absorbed powers and were easily
overcome by implementing a simple bisection procedure, averaging the
perturbation of successive iterations (see figure
\ref{fig:Convergence}).

These convergence issues are numerical rather than physical. They
arise from our instantaneous approach to a system that exhibits
thermal lag. Our model treats the thermal response of the mirror as
being comparably rapid to the optical buildup within the cavity.  In
reality the cavity response is many orders of magnitude faster, and
the optical field adapts nearly instantaneously to the thermal
deformation of the mirror, but not vice-versa.

Nevertheless we believe that our result represents the true physical
solution. If we reduce our `time step size' adding only a small
portion of the mirror perturbation (reality being the limit of
infinitesimal step size) we arrive at the same equilibrium, but with
much slower convergence.

\section{Implications for interferometric detectors}

\subsection{Thermoelastic deformation and resultant mode shape}
\label{sec:DeformationAndModeShape}

\begin{figure}[htbp!]
\centering
\includegraphics[width=0.49\textwidth]{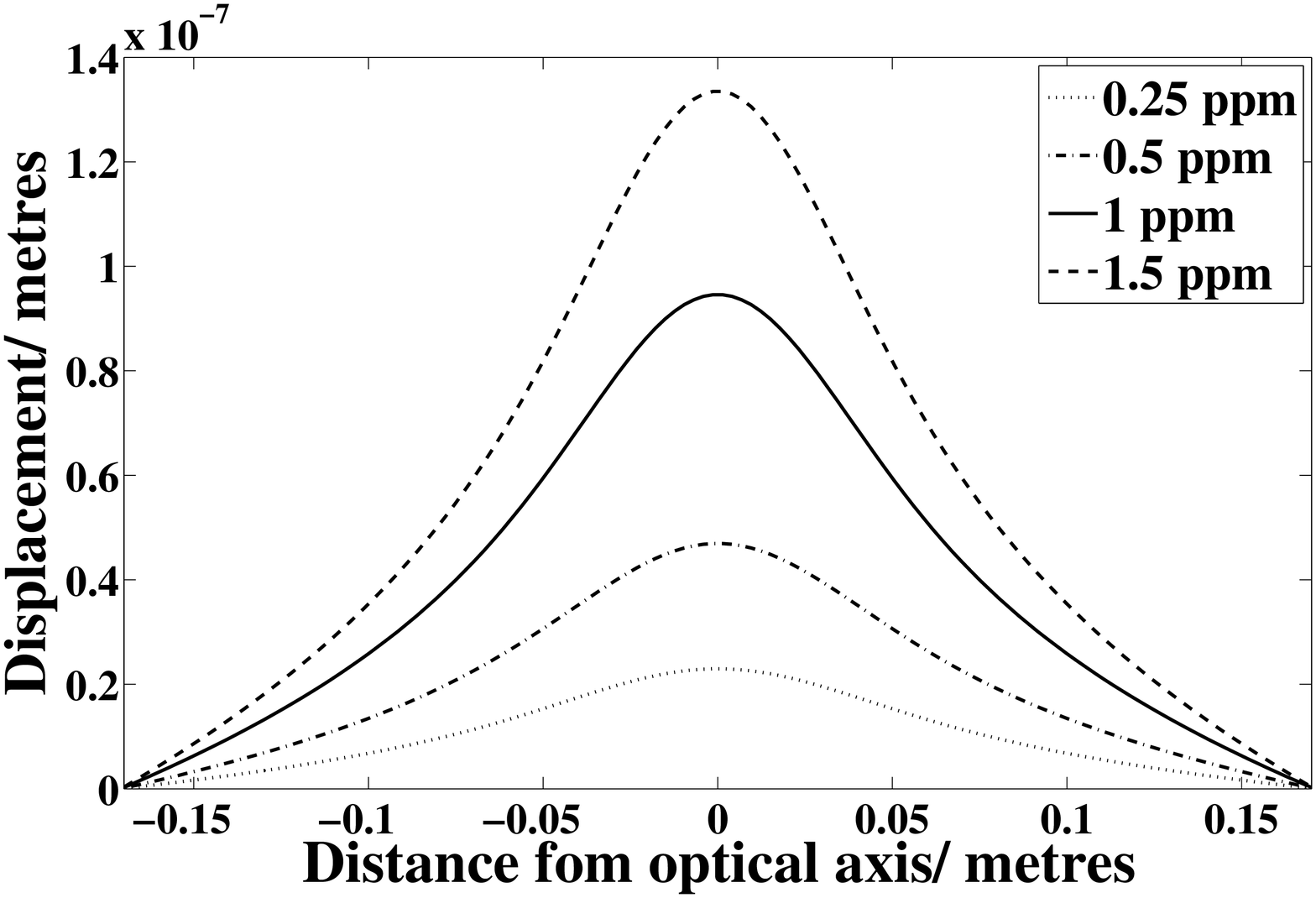}
\includegraphics[width=0.49\textwidth]{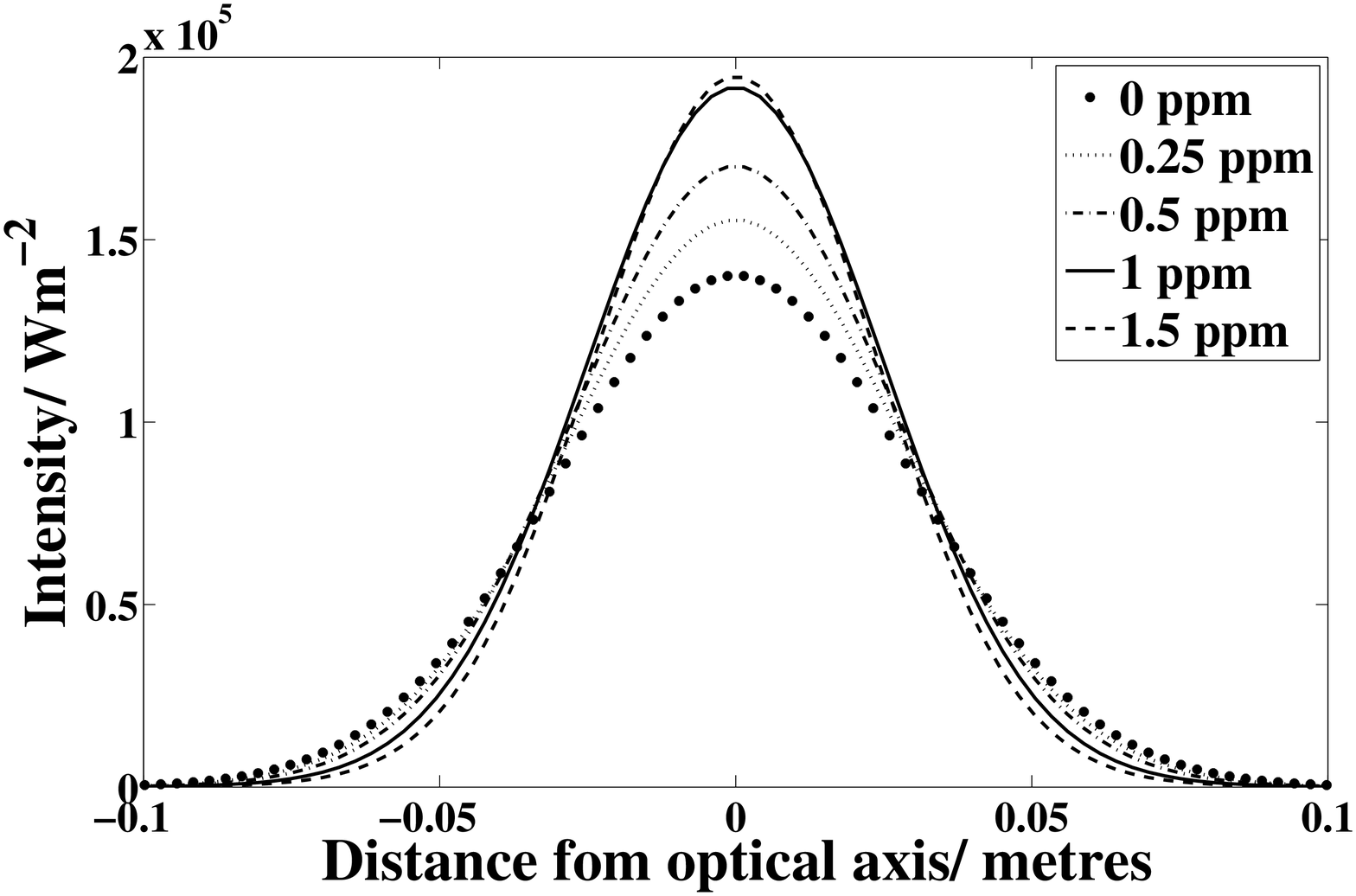}\\
\includegraphics[width=0.49\textwidth]{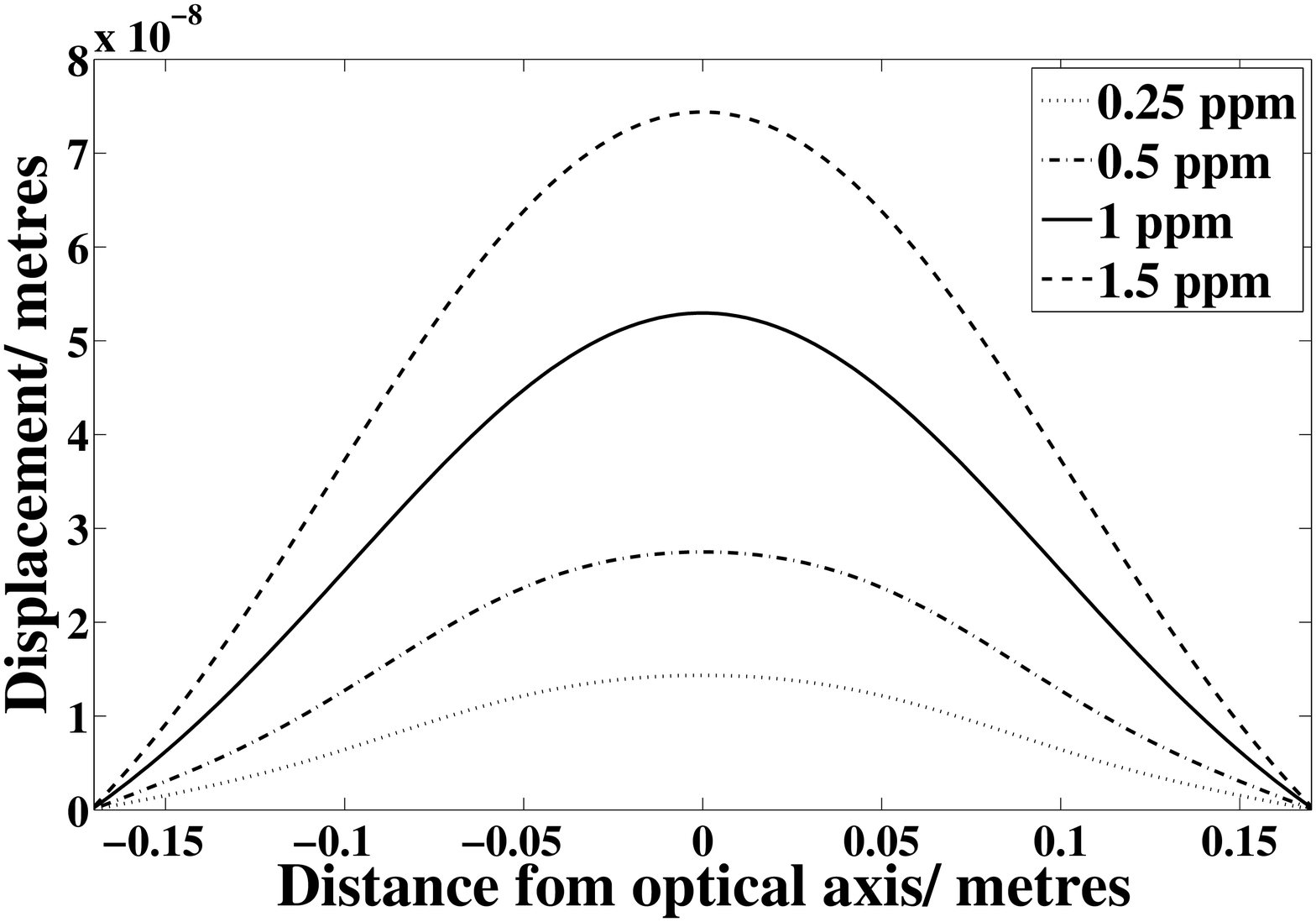}
\includegraphics[width=0.49\textwidth]{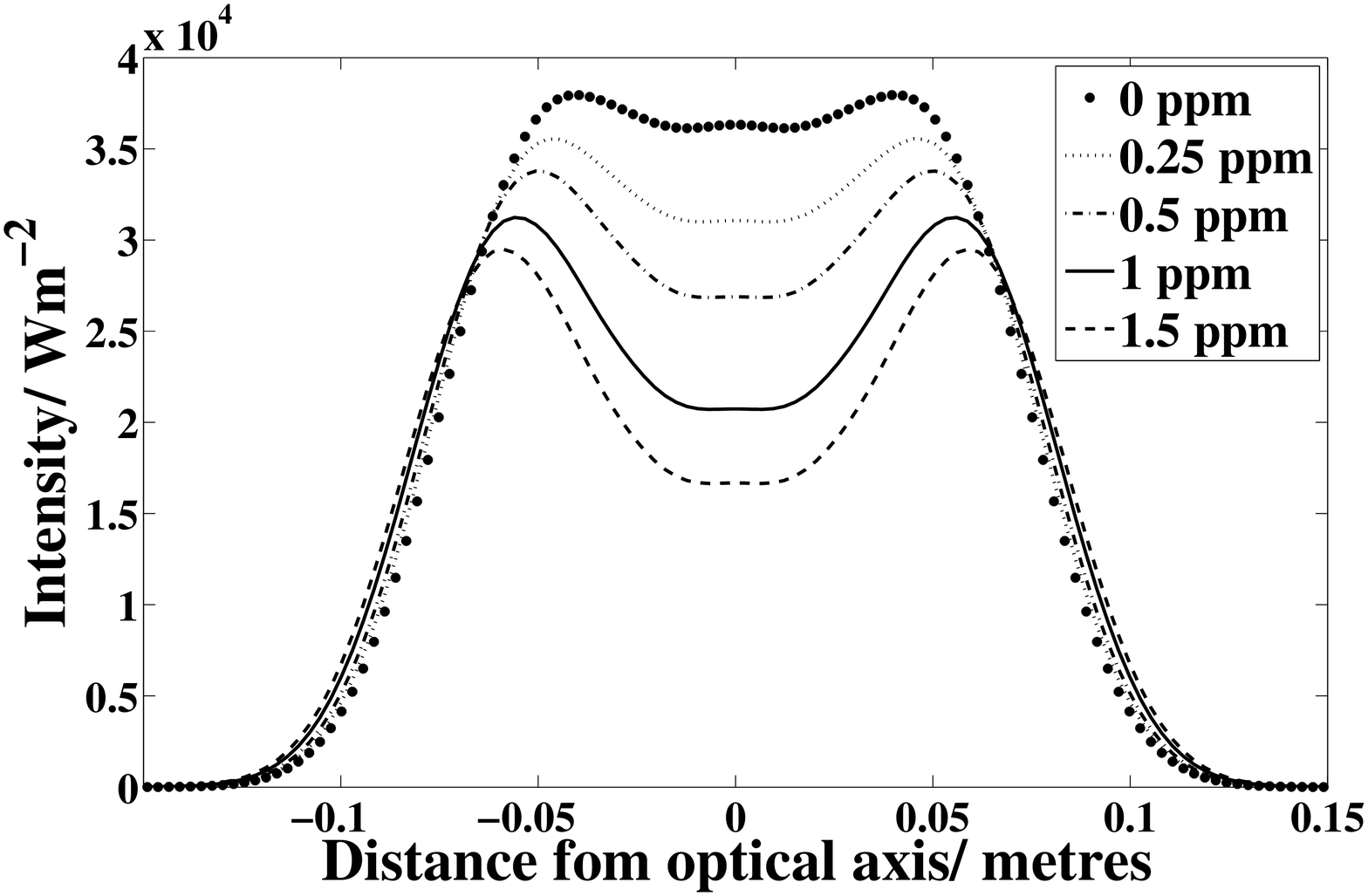}\\
\includegraphics[width=0.49\textwidth]{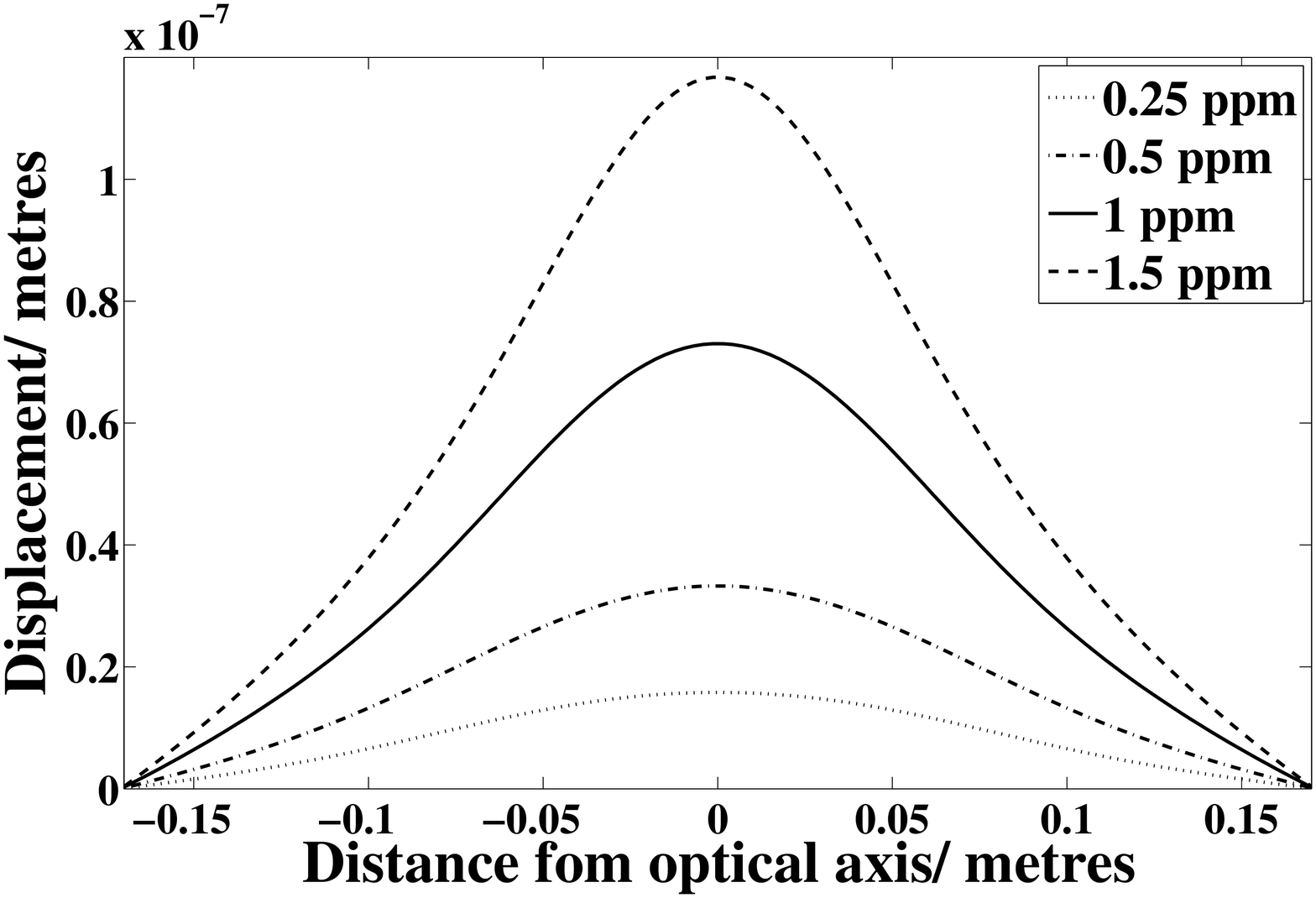}
\includegraphics[width=0.49\textwidth]{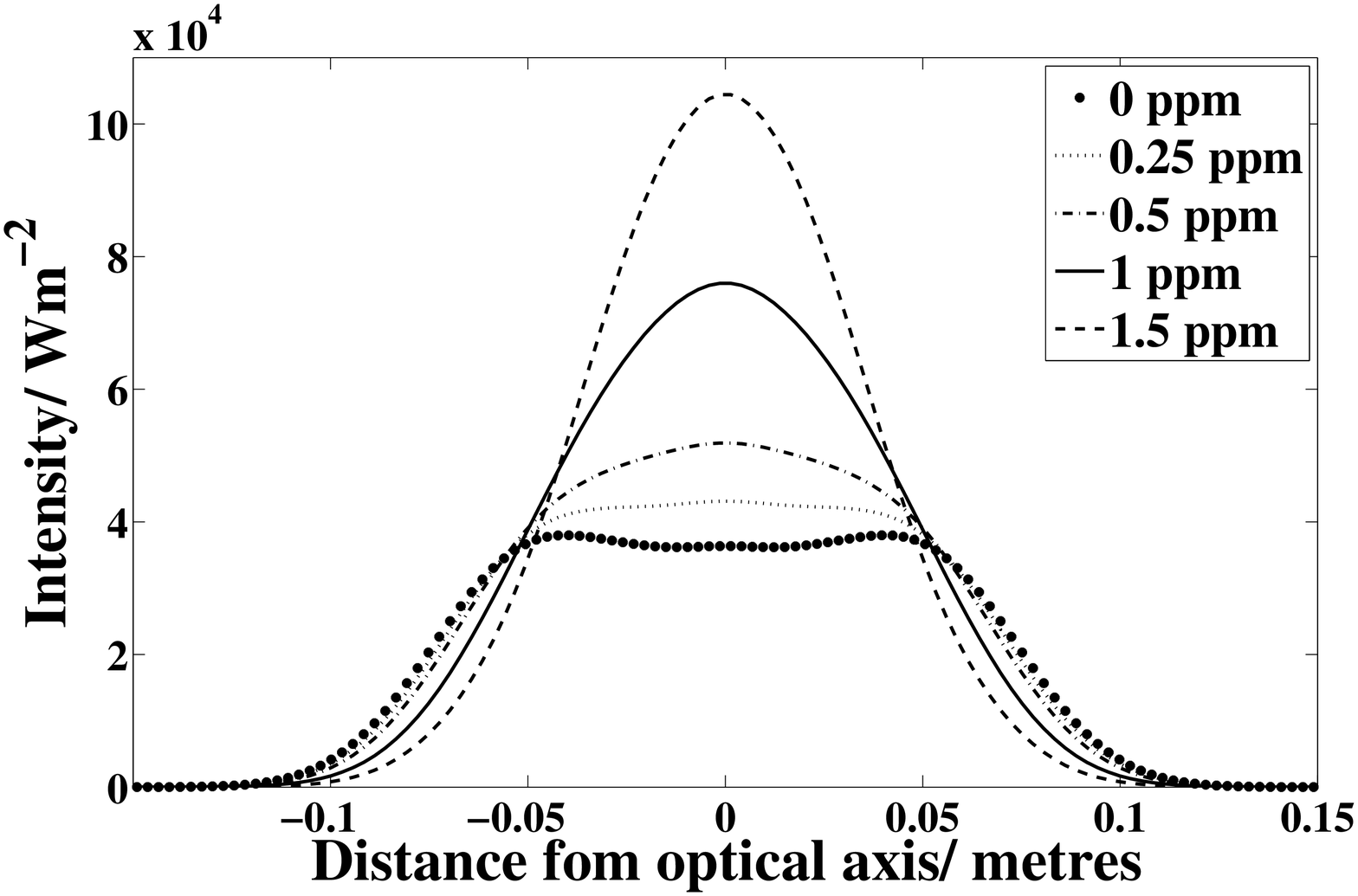}\\
\caption{Thermoelastic deformation (left) and resonant mode shape
  (right) as a function of coating absorption. To ease extrapolation,
  intensities are plotted for 1~W of incident laser power. Top row:
  spherical cavity; middle row: flat mesa cavity; bottom row:
  concentric mesa cavity.}
\label{fig:TEdeformation}
\end{figure}

The leftmost column of figure \ref{fig:TEdeformation} shows the
deformation of the mirror profile for both mesa and Gaussian
cavities. The shape of the deformation is dependent on the thermal
gradients imposed by the beam.
At low power levels flat and concentric mesa beams induce similar
thermoelastic deformations; as greater power is absorbed the
concentric beam tends toward a Gaussian intensity profile and hence
gives rise to larger deformations typical of Gaussian beams. Flat
mesa beams retain their greater width under thermal perturbations
and produce about 50\% less thermal deformation than that produced
by the Gaussian mode.  This is consistent with the general results
of Vinet \cite{Vinet90}, and is due to the more even deposition of
heat into the mirror.  The mesa deformation more closely resembles a
pure radius of curvature change, which can be easily compensated by
heating the rear face of the mirror, an approach clearly better
suited to the end test masses of gravitational wave detectors.

Although the shapes of the thermal distortions are of interest the
change in the structure of the resonant light field is more important
to the performance of the interferometer. The right-hand column of
figure \ref{fig:TEdeformation} shows the effects of the deformations
on the cavity eigenmode.  The Gaussian beam is fairly robust in its
functional form as the absorbed power increases.  For small amounts of
heating, the spot size on the mirrors decreases, as the thermoelastic
bump effectively increases the mirrors' radii of curvature, making the
cavity more stable.  The stored power, as we show in the next section,
does not substantially decrease until the absorbed power becomes
relatively large.

The mesa beam cavities, on the other hand, undergo striking changes.
The flat mesa beams deform into a more annular shape, even for the
smallest amounts of heating, whilst the width of the profile changes very
little.  This is likely to be due to confinement of the optical field by the
steep rim of the mirror profile, which is not much changed by the
thermal distortion. The concentric mesa beams are also grossly
deformed but instead of retaining their width these beams become
strongly peaked.

This differing behaviour may be understood by considering the profiles
of the two mesa mirrors (see figure \ref{fig:Comparison}). Both
mirrors are only a small positive deviation away from optics which
support narrow Gaussian modes. The flat mesa mirror is realised by
\emph{adding} a small deviation, $z_{\mathrm{HR}}$, to a flat surface;
the concentric mesa mirror is constructed by \emph{subtracting} the
same small deviation from a spherical surface. Thermal effects
\emph{add} a small perturbation to the existing mirror profile making
the flat mesa mirror even less like a flat surface and the concentric
mesa mirror more like a spherical surface. Hence these effects will
tend to push the concentric mesa mirror toward supporting narrower
beams whilst the flat mesa mirror should be more resilient.

\subsection{Losses}
Given the high finesse of the arm cavities, losses are significant
even at the part per million level. In table \ref{Tab:Losses} we
present a summary of the diffraction and mode matching losses as a
function of absorbed power.
\begin{table}[htbp!]
\caption{Cavity gain and diffraction losses as a function of coating
  absorption. The quoted diffraction loss is for a complete round trip,
  losses per bounce are half as large. Mode matching losses refer to the
  fraction of the input beam not coupled into the cavity and are
  derived from the discrepancy between the theoretical and observed
  intracavity powers, accounting for diffraction losses but not other
  sources of intracavity loss (e.g. absorption).} \label{Tab:Losses}
\lineup
\begin{indented}
\item[]
\begin{tabular}{@{}lllll}
\br
Cavity      & Coating            & Cavity      & Diffraction           & Mode Matching\\
            & Absorption         & Gain        & Loss                  & Loss\\
            & ppm                &             & ppm                   & \%\\
\mr
            & 0                  & 795         & \0\00.43              & 0\\
            & 0.25               & 792         & \0\00.19              & $<1$ \\
Gaussian    & 0.5                & 786         & \0\01.16              & 1 \\
            & 1                  & 755         & \032.42               & 4 \\
            & 1.5                & 689         & 189.32                & 7 \\
\mr
            & 0                  & 755         & \0\00.48              & 5\\
            & 0.25               & 747         & \0\00.80              & 6 \\
Flat        & 0.5                & 737         & \0\01.37              & 7 \\
Mesa        & 1                  & 717         & \0\03.20              & 10 \\
            & 1.5                & 697         & \0\06.63              & 12 \\
\mr
            & 0                  & 756         & \0\00.49              & 5\\
            & 0.25               & 763         & \0\00.33              & 4 \\
Concentric  & 0.5                & 768         & \0\00.29              & 3 \\
Mesa        & 1                  & 764         & \0\00.76              & 4 \\
            & 1.5                & 733         & \0\06.12              & 8 \\
\br
\end{tabular}
\end{indented}
\end{table}

This increased mode matching loss effectively reduces the
cavity gain. The mode matching loss is also significant for any
heterodyne readout scheme employing phase modulated sidebands of
Gaussian profile.

The altered cavity eigenmode will not couple as strongly to the input
beam. We derive this dominant mode coupling loss from a comparison between
the theoretical intracavity power and that which is seen in SIS. Once
diffraction effects are taken into account we attribute the residual
losses to mode coupling error, in doing so we ignore other effects
such as scatter and absorption. The numbers obtained using this method
are in excellent agreement with those calculated directly from the
inner product of the intracavity and injected fields. One could
envisage mitigating these losses via thermal compensation in the
recycling cavity. Such ideas are not discussed in this article.

Our calculated round trip diffraction losses for the unperturbed
cavities are in accord with previously published
values \cite{O'Shaughnessy04,BhawalTech05}. To our knowledge the
results for perturbed cavities are the first to be published.

\subsection{Thermal noise}
Non-Gaussian beams, including mesa beams, are being studied both
theoretically and experimentally as they are expected reduce test mass
thermal noise in interferometric gravitational wave detectors
\cite{Tarallo07,Bondarescu06,Mours06,BondarescuThesis}.  Thermoelastic
distortion of the cavity mirrors changes the intensity profile of the
cavity mode profile and thus alters the effects of thermal
noise.\footnote{Strictly speaking, thermoelastic distortion is
  associated with increased thermal noise by virtue of the extra heat
  in the mirror.  We ignore this effect in this article.}

It has been shown that a thermally perturbed spherical cavity
continues to support a nearly Gaussian beam \cite{SmithTech07}. The
only consequence of moderate heating is that the beam waist shrinks,
increasing total thermal noise by around 10\%.

\begin{figure}[htbp!]
\begin{minipage}{\textwidth}
\centering
\includegraphics[width=\textwidth]{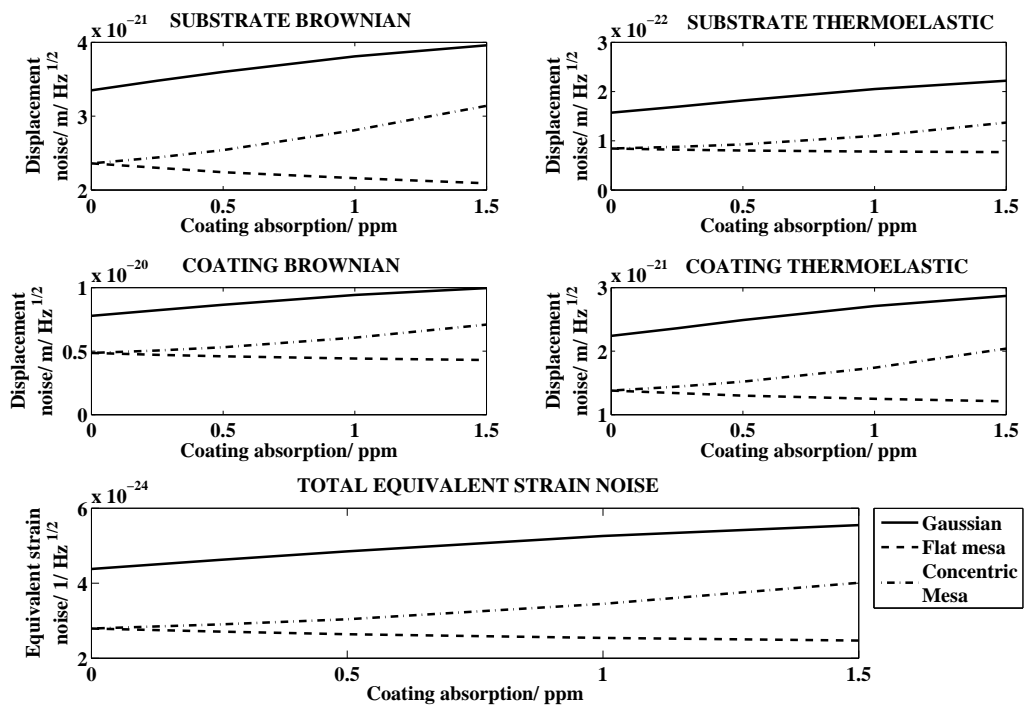}
\caption{Thermal noise as a function of coating absorption. All values
  are evaluated at 100 Hz for a cylindrical fused silica substrate
  (34x20 cm) with a silica-tantala quarter wave coating. A full list
  of physical parameters is given in \ref{sec:MaterialParameters}. We
  find that the impact of thermal noise associated with mesa beams
  decreases as a function of coating absorption for flat
  configurations and increases in the concentric scheme. Total
  equivalent strain noise is evaluated as
  $\sqrt{4\sum_{i}N_i^2}/L$. Each $N_i$ represents the displacement
  noise arising from one of the four mechanisms plotted above and $L$
  is the length of the arm cavity. This calculation assumes that all
  four cavity optics have the same coating as the highly reflective
  end mirrors. In reality the input mirrors are less reflective, have
  thinner coatings and therefore exhibit lower thermal noise. Hence
  our calculation overestimates the total thermal noise.}
\label{Fig:Noise}
\end{minipage}
\end{figure}

Thermal effects in non-Gaussian cavities are less well understood.
Using the techniques outlined in \ref{sec:ThermalNoise} we
calculated the thermal noise expected for the perturbed eigenmodes
of section \ref{sec:DeformationAndModeShape}.  Our findings are
presented in figure \ref{Fig:Noise} along with the corresponding
results for a Gaussian cavity.

As expected the thermal noise associated with the Gaussian and
concentric mesa beams increased with absorbed power as the beam waist
became smaller. The effects of beam size on thermal noise have been
well documented
\cite{BHV98,Braginsky99,Harry02,Braginsky03,Fejer04}. However, the
noise of the thermally perturbed flat mesa beam decreased under the
same conditions. Note that these results are a strong function of the
material parameters used in their evaluation, see
\ref{sec:MaterialParameters}.


\section{Thermal Compensation System}

Although the above improvements in the thermal noise seen by a flat
mesa beam are interesting, the modes which produce them would
simultaneously reduce the sensitivity of any detector as they have a
poor overlap with the Gaussian modes outside the interferometer's
arms. We must maintain the standard mesa mode even if the thermal
noise will be greater\footnote{It is of course possible to devise a
  scheme whereby gravitational wave readout is effected by injecting a
  suitable mode at the output port of the interferometer.}.

The Mexican hat mirrors which support mesa beams are constructed using
a multi-step silica deposition process over a micro-polished flat
substrate \cite{SimoniThesis}. Currently this technique can achieve up
to 2 nm precision and is able to create almost any mirror profile
desirable in a full scale interferometric detector. Exploiting this
technology we resolved to design a mirror which only achieves the
correct profile after thermoelastic deformation caused by the impinging
optical power. This approach would reduce the compensation required
from (and hence noise introduced by) external sources such as a carbon
dioxide laser or ring heater.

\subsection{Method}
Using the tools developed above we are able to find the
thermoelastic deformation caused by an (almost) arbitrary intensity
profile.  Using SIS we can find the eigenmode of a thermally
perturbed cavity. The self-correcting mirror profile, giving the
desired cavity eigenmode only when thermally deformed, may then be
found iteratively. Beginning from an unperturbed cavity, the system
is allowed to evolve to its steady state as described by figure
\ref{fig:Flowchart}. We then subtract the resulting thermoelastic
deformation from the nominal mirror profiles and allow the system to
reach a new steady state. Iteratively repeating this process one
eventually arrives at the mirror profile which deforms under thermal
loading to support the desired mode.

\subsection{Results}

In figure \ref{fig:TCS} below we show how such a system might operate for
Gaussian and mesa modes. We chose to study the case of 0.5 ppm coating
absorption, at the time of writing a typical value for future
gravitational wave detectors.

\subsubsection{Gaussian}
\begin{figure}[htbp!]
\centering
\includegraphics[width=0.49\textwidth]{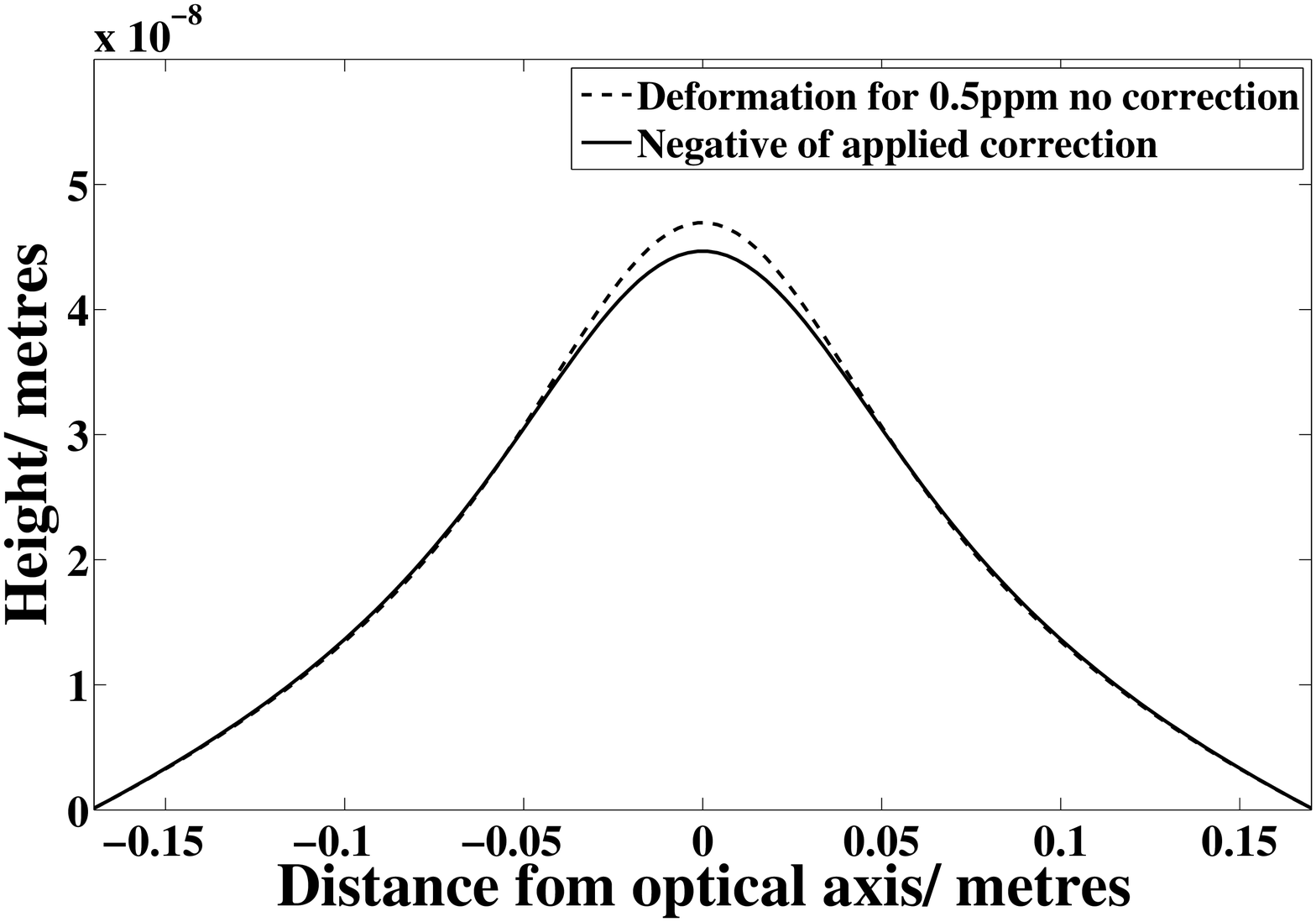}
\includegraphics[width=0.49\textwidth]{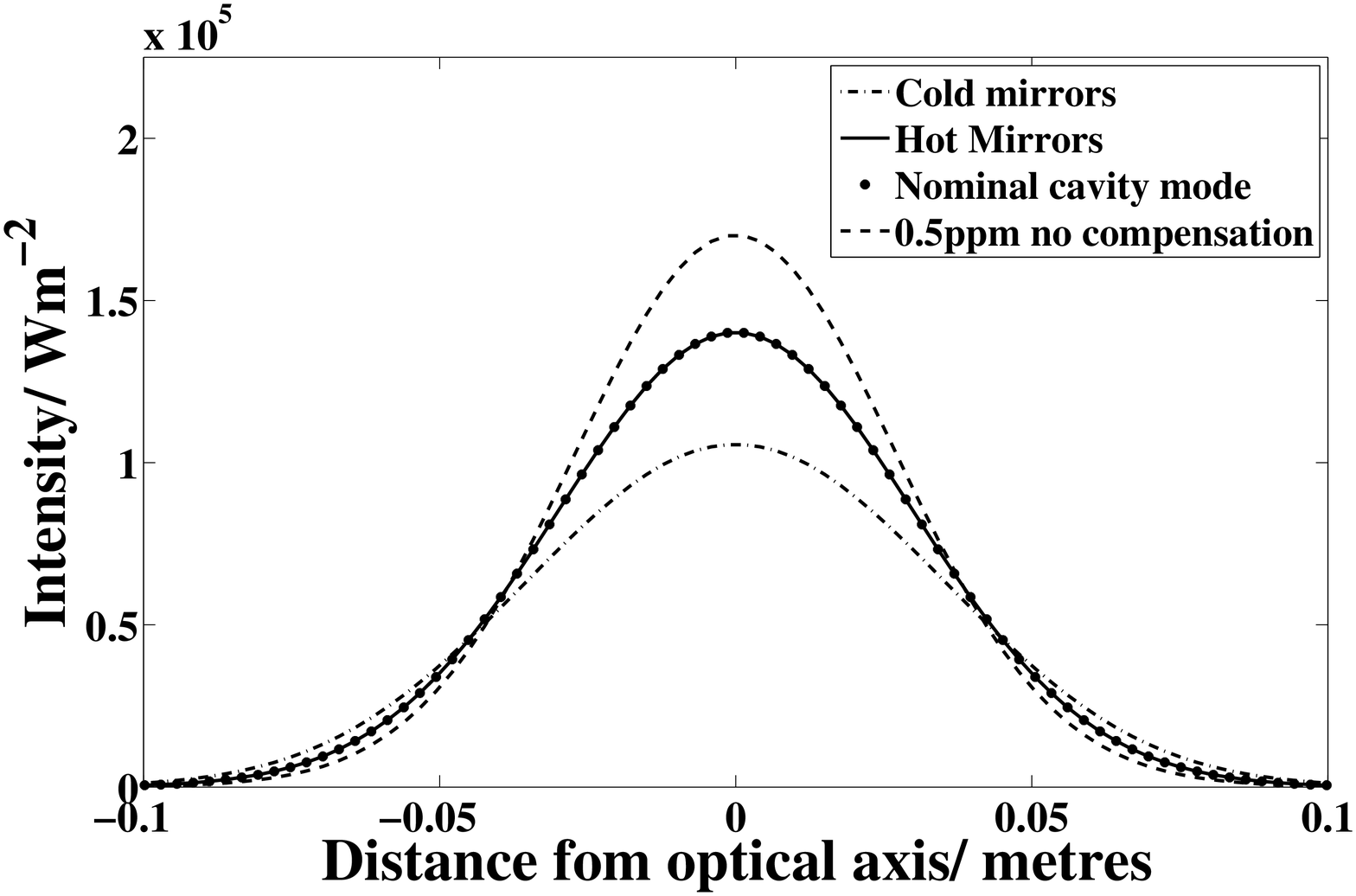}\\
\includegraphics[width=0.49\textwidth]{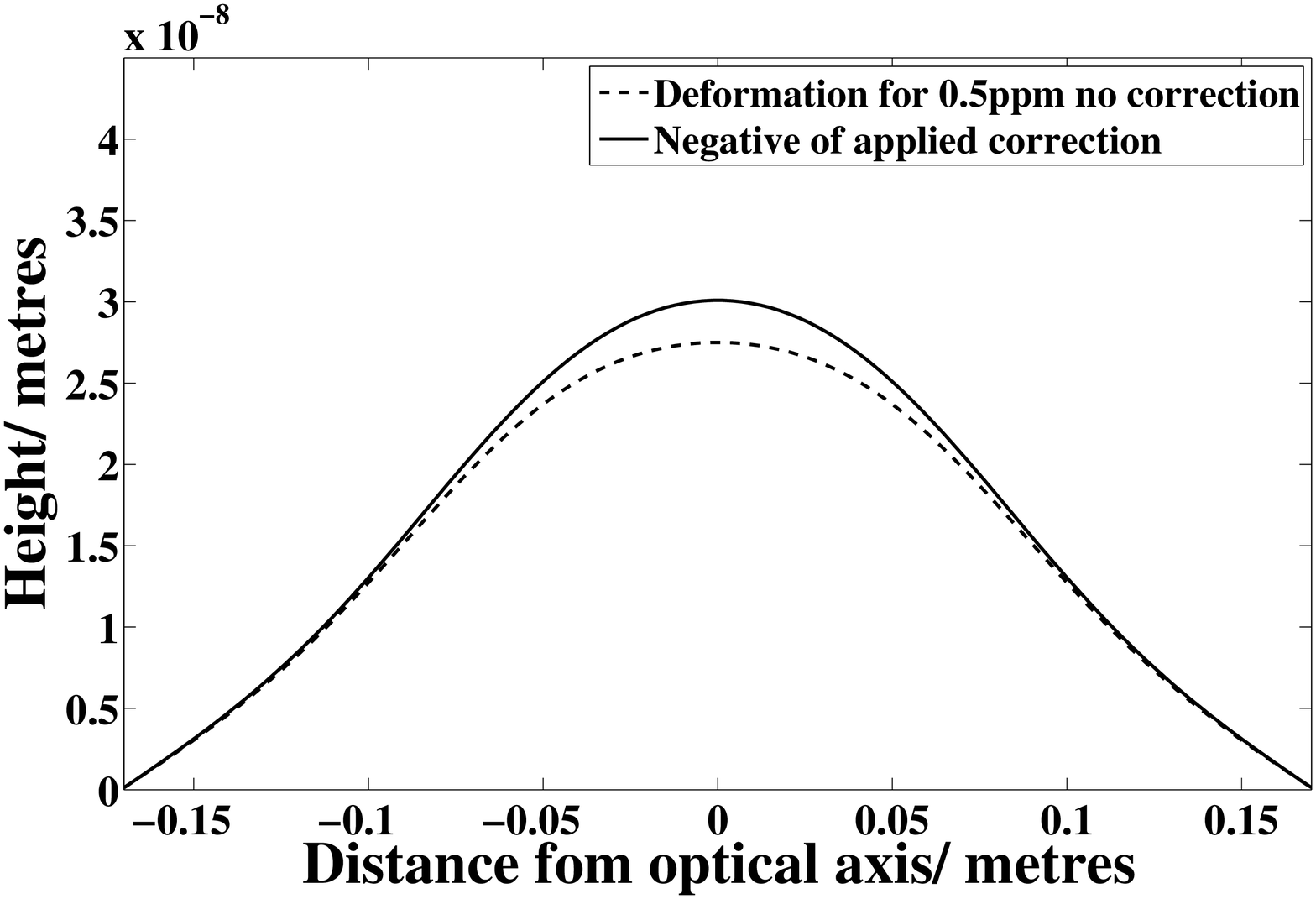}
\includegraphics[width=0.49\textwidth]{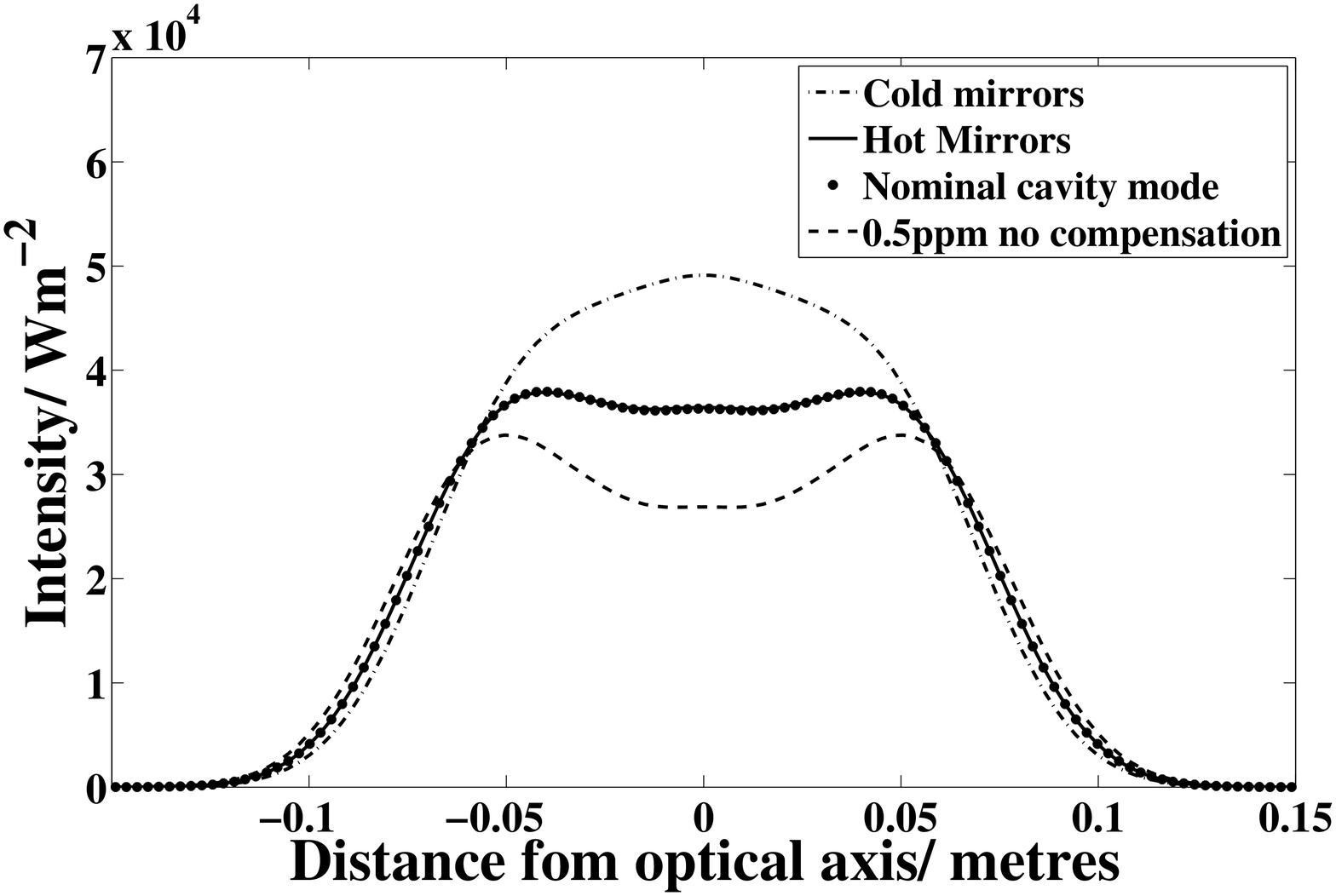}\\
\includegraphics[width=0.49\textwidth]{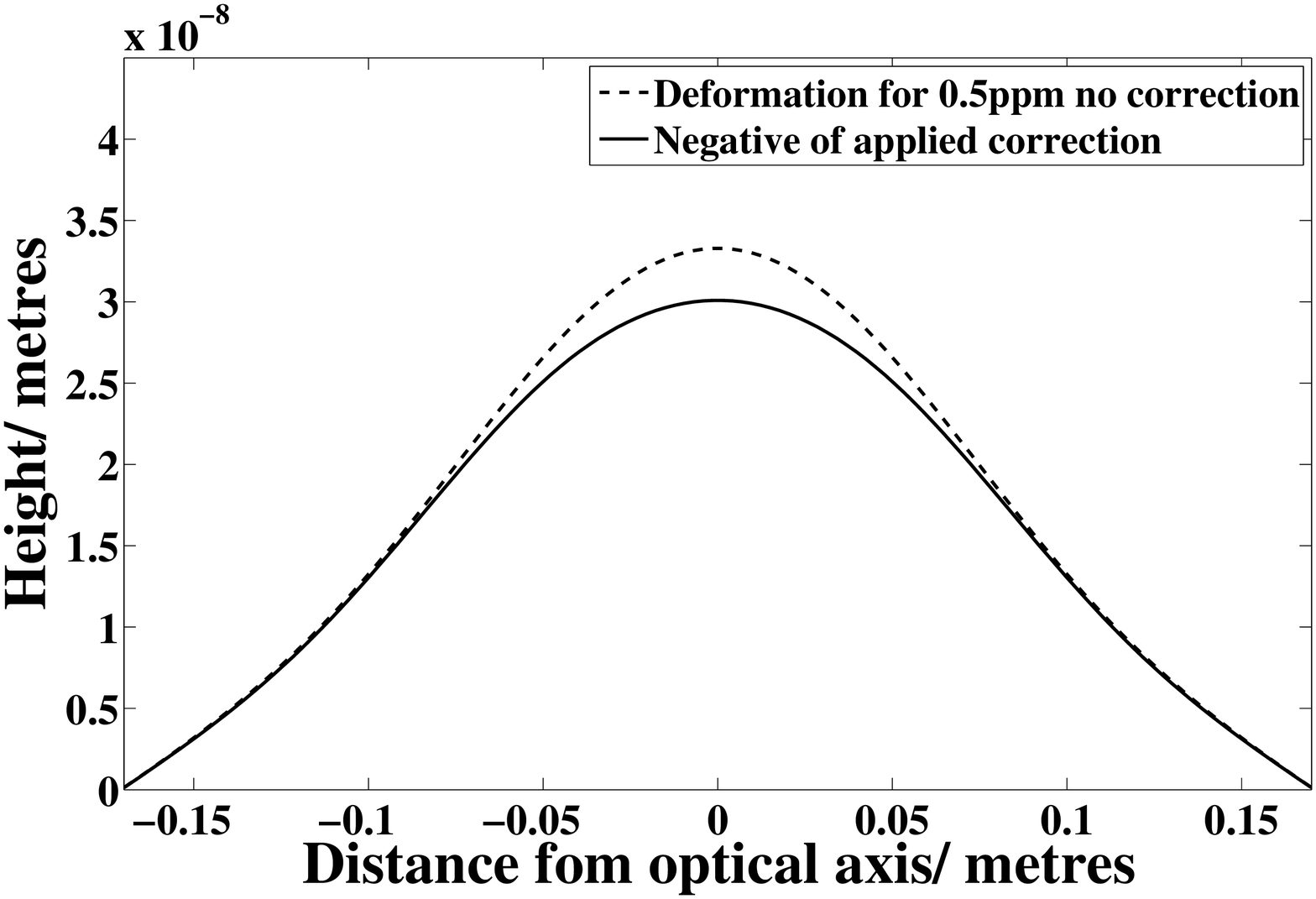}
\includegraphics[width=0.49\textwidth]{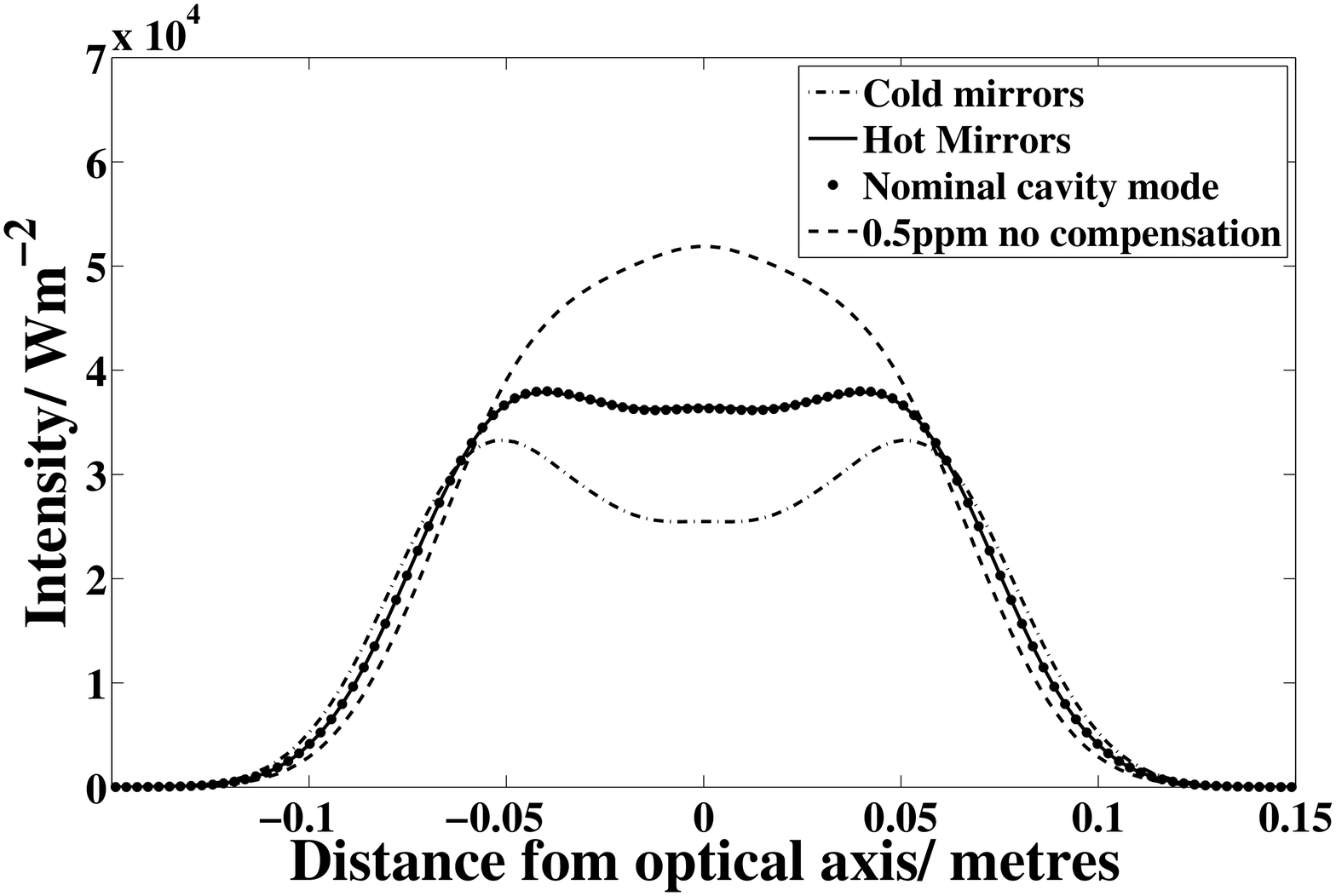}\\
\caption{Thermal compensation idea: Gaussian (top row), flat mesa
  (middle row), concentric mesa (bottom row). Left: Thermoelastic
  deformation with no correction and the correction which must be
  subtracted to regain the nominal mode. Right: Mode profiles with no
  TCS (dashed), with our doctored TCS mirrors `cold' (dash-dot) and
  `hot' (solid). Note that the recovered mode (solid line) overlaps
  exactly with the ideal cavity mode (filled circles). Again
  intensities are plotted for 1~W of incident laser power.}
\label{fig:TCS}
\end{figure}

The upper-left plot of figure \ref{fig:TCS} shows the uncorrected
deformation arising from 0.5 ppm absorption (dashed line). The solid line is
the profile which must be subtracted from our cavity mirrors to
restore the nominal mode. Note that these profiles are not
equal. Qualitative understanding of this effect may be gained by
noting that the power stored in the cavity formed by the compensating
mirrors when thermal absorption is absent (upper right plot, dash-dot curve) is less
than that present in the deformed cavity (upper right plot, dashed). Since less power
is incident on the optics, the thermoelastic deformation induced is
smaller and hence a smaller correction is necessary to restore the
nominal cavity eigenmode.\footnote{An identical argument holds for the
  concentric mesa beam case whilst an analogous approach is suitable
  for flat mesa beams where the cold cavity stores more power and
  requires a correction larger than the uncorrected deformation.}

In the upper-right plot we outline our compensation scheme: the
circular markers give the theoretical cavity eigenmode ignoring
thermal effects, the dashed line represents the intensity profile to
be expected if no correction is implemented. Using a dash-dot line we
show the mode which is resonant when our compensating mirrors are
cold. The solid line shows the intensity profile recovered once these
mirrors are at operating temperature, as expected this profile
agrees excellently with the ideal cavity mode.

\subsubsection{Mesa}
The middle and bottom rows of figure \ref{fig:TCS} show analogous
results for flat mesa and concentric mesa cavities respectively. For
both configurations the mode recovered after heating again shows superb
agreement with the nominal mode.

In order for the corrective mirror profiles calculated above to be
practicable in a real interferometer we may require some auxiliary
source to heat the test mass before resonance is attained (such as a
carbon dioxide laser or ring heater). Once stably locked this
compensating source may have its heating significantly reduced so that
noise is injected at a level which is acceptable for recording
astrophysical data.

The results of both of these figures neglect multiple real-world
effects. For example vague knowledge or variability in the coating
absorption, laser intensity noise and fabrication errors. Effects such
as these were responsible for the variable success of a similar scheme
used in the polishing of the initial LIGO power recycling
mirrors. Nonetheless we believe that this approach merits further
study.

\section{Summary and discussion}
We have calculated the change in the resonant mode of a mesa beam
Fabry-Perot cavity as a function of coating absorption. Along with
other candidates, this non-Gaussian beam is being considered as an
option for future interferometric gravitational wave detectors as it
ameliorates the effects of test mass thermal noise. We find for flat
mesa beams, in contrast to Gaussian and concentric mesa modes, that
the thermal noise mitigation increases with absorbed power.

In addition we have outlined a possible passive method of thermal
compensation for non-Gaussian beams. The same techniques may also be
useful as an addition to TCS systems for Gaussian beams.

\ack We wish to thank Jerome Degallaix, Andri Gretarsson, Eric
Gustafson, Norna Robertson, Sheila Rowan, Ken Strain and Jean-Yves
Vinet for valuable assistance throughout this work.  The authors
acknowledge the support of the University of Glasgow, The Carnegie
Trust for the Universities of Scotland and the LIGO Laboratory. LIGO
was constructed by the California Institute of Technology and
Massachusetts Institute of Technology with funding from the National
Science Foundation, and operates under cooperative agreement
PHY-0107417. This paper has LIGO document LIGO-P080063-00-Z.


\appendix

\section{Thermoelastic deformation}
\label{sec:ThermoelasticDeformation}

We follow a well-known derivation by Hello and Vinet \cite{Vinet90} and recent
expansion by Vinet \cite{Vinet07} which allows one to calculate the
thermoelastic deformation induced in any axially symmetric
mirror heated by absorption of an axially symmetric transmitted beam.

This derivation solves for the displacement of the mirror coating along the beam axis
$u_z(r)$ in terms of a Dini expansion:
\begin{equation}
u_z(r)=\sum_mU_m(1-J_0(\zeta_mr/a))-\frac{1-\nu}{2Y}Br^2
\end{equation}
where $\zeta_m$ is the $m^{\mathrm{th}}$ root of $\zeta
J_1(\zeta)-\tau J_0(\zeta)=0$, $\tau=4\sigma'T^3a/K_{\mathrm{th}}$
being the reduced radiation constant and $a$ the radius of the
mirror. Here $T$ is absolute temperature, $K_{\mathrm{th}}$ is thermal
conductivity and $\sigma'$ is the Stefan-Boltzmann constant corrected
for emissivity. The coefficients $U_m$ are given by
\begin{eqnarray}
\fl &U_m&=\frac{\alpha(1+\nu)\epsilon
P_{\mathrm{L}}a^2}{K_{\mathrm{th}}}\sum_m\frac{p_m}{\zeta_m}\frac{\zeta_m+\chi-(\zeta_m-\chi)\exp(-2\zeta_m h/a)}{(\zeta_m+\chi)^2-(\zeta_m-\chi)^2\exp(-2\zeta_m h/a)}\\
\fl \mathrm{where}\quad &p_m &=
\frac{1}{P_{\mathrm{L}}}\frac{2\zeta_m^2}{a^2(\chi^2+\zeta_m^2)J_0^2(\zeta_m)}\int_0^a|\Psi(r)|^2J_0(\zeta_m
r/a)r\mathrm{d}r
\end{eqnarray}
$P_{\mathrm{L}}$ is the power circulating in the cavity and $\epsilon$ is a
coating absorption rate so that the product $P_{\mathrm{L}}\epsilon$ gives the
total absorbed power in the coating, $h$ is the thickness of the
mirror, and $\alpha$ and $\nu$ are the mirror thermal expansion
coefficient and Poisson's ratio.  The second term of $u_z(r)$ is the
Saint-Venant term, in which $Y$ is the mirror Young's modulus.  The
calculation of $B$ is given by Hello and
Vinet \cite{Vinet90}.\footnote{The sign of the Saint-Venant correction
  is incorrect in Hello and Vinet's final result, and has been
  corrected here.}  This term acts to make the thermal deformation
more convex.

\begin{figure}[htbp!]
\centering
\includegraphics[width=0.49\textwidth]{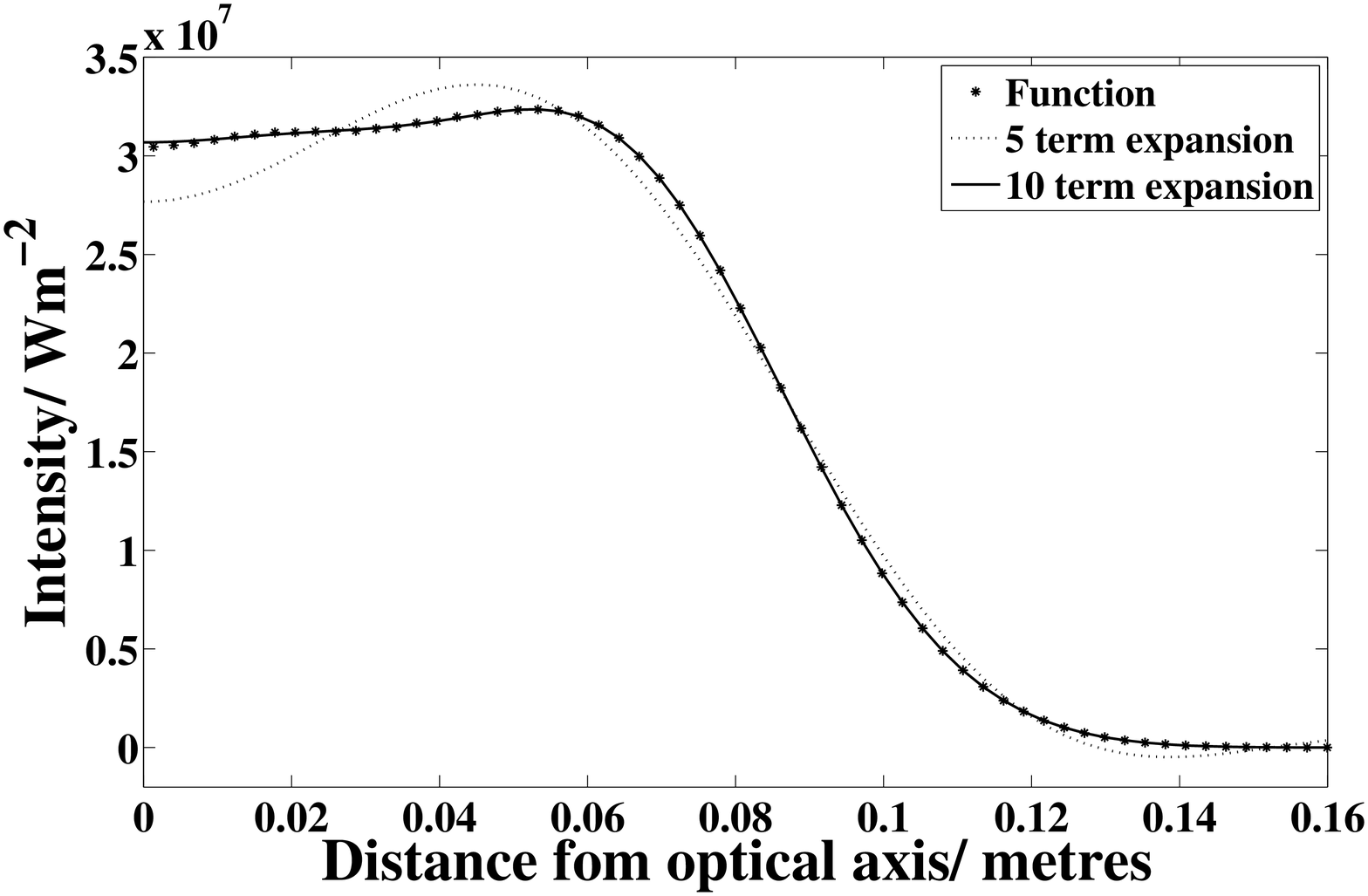}
\includegraphics[width=0.49\textwidth]{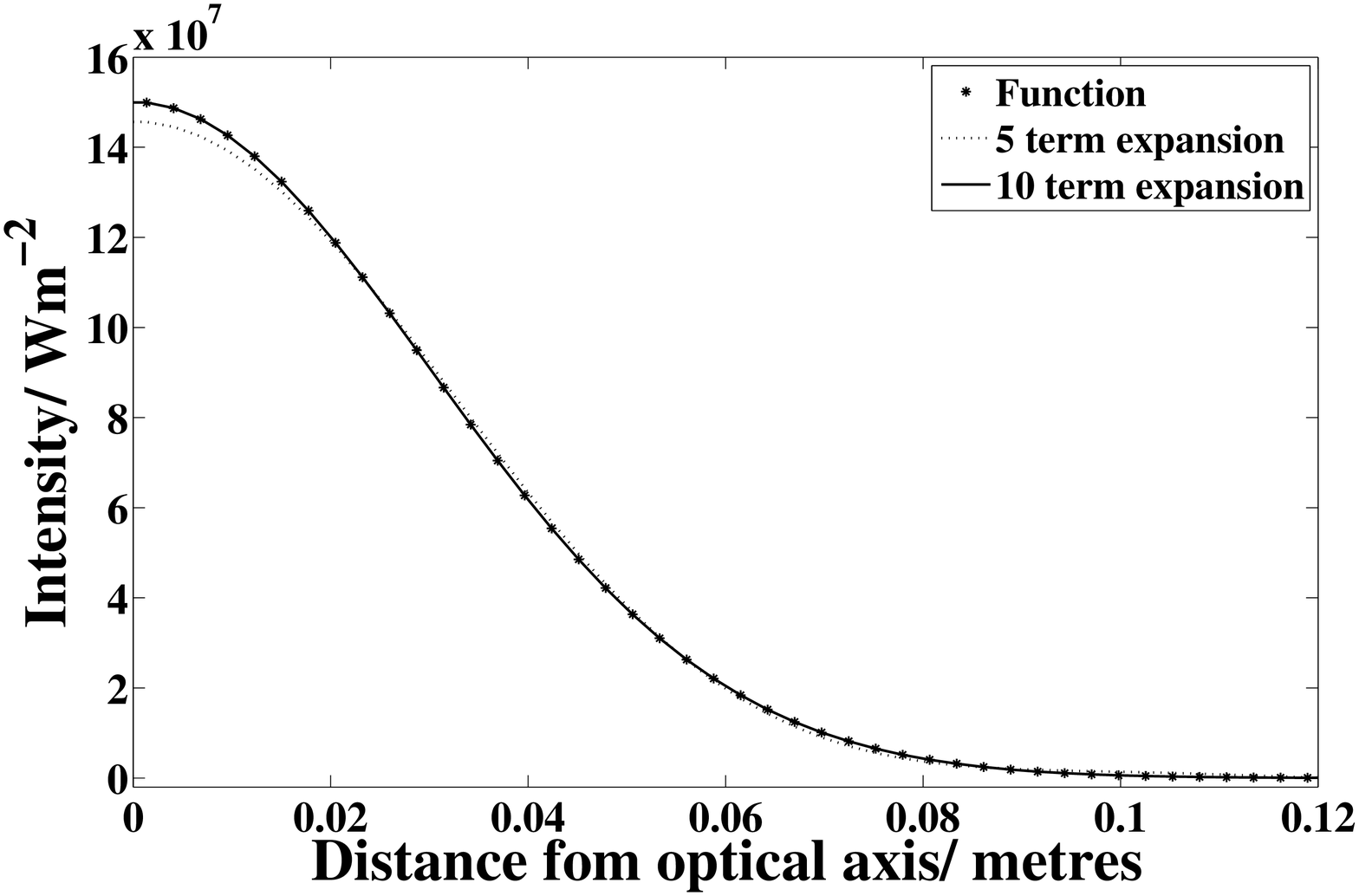}
\caption{Much of the work described herein relies on Dini
  expansions. Here we show how efficiently the Dini expansion is able
  to approximate our two beams under study. Beams are normalised to
  850~kW of integrated power. Left: A recreation of the mesa beam
  (asterisks) using 5 and 10 terms. Right: An analogous plot for
  Gaussian beams. The functions are well reconstructed with a minimum
  of terms.}
\label{fig:DiniExpansions}
\end{figure}

Figure \ref{fig:DiniExpansions} shows how efficiently the Dini
expansion reconstructs both Gaussian and mesa profiles. Excellent
fidelity is achieved with few terms. Our analysis is adaptive,
unconditionally implementing the first ten terms with subsequent terms
being added if they are within a factor of $10^{-6}$ of the principal
term. For the unperturbed beams this corresponds to over 80 terms.

\section{Thermal noise}
\label{sec:ThermalNoise}
In \cite{Levin98} Levin takes a fluctuation-dissipation theorem approach to
thermal noise calculation, we follow his example. The spectral
density of displacement noise due to thermal effects is given by
\begin{equation}
S_X(\omega)=\frac{8k_{\mathrm{B}}T}{\omega^2}\frac{W_{\mathrm{diss}}}{F_0^2}
\label{NoisePSD}
\end{equation}
where $W_{\mathrm{diss}}$ is the average energy dissipated in the region of
interest (coating or substrate) in response to an applied oscillatory
pressure
\[P(\vec{r},t)=\Re(F_0f(\vec{r})\exp(\rmi\omega t))\]
which has the same spatial distribution, $f(\vec{r})$, as the intensity of the
beam under study. Our
task is to calculate $W_{\mathrm{diss}}$ for each of the dissipative
mechanisms in which we are interested.

In this section, to maintain agreement with previous publications, the test
mass occupies the region $r\in[0,a]$, $z\in[-h/2,0]$, with the coated
surface at $z=0$.

\subsection{Stresses and strains}
\label{sec:stressstrain}
In order to calculate $W_{\mathrm{diss}}$ - and therefore the thermal noise
associated with our perturbed modes - we must first find the stresses
and strains in the substrate and coating. Here we adopt the techniques
of BHV \cite{Bondu98} subsequently corrected by Liu \&
Thorne \cite{Liu00}.

Again we form a solution in terms of Dini expansions, seeking
displacements of form:
\begin{eqnarray}
u_r(r,z)&=\sum_mA_m(z)J_1(k_mr)\\
u_z(r,z)&=\sum_mB_m(z)J_0(k_mr)\\
u_{\phi}(r,z)&=0
\end{eqnarray}

The calculation of the $A_m$ and $B_m$ follows (33-35) of
Liu and Thorne \cite{Liu00}. To take account of the intensity profile, $|\Psi(r)|^2$,
of our thermally perturbed beams (36) takes the form,
\begin{equation}
p_m =
\frac{1}{P_{\mathrm{L}}}\frac{2}{a^2J_0^2(\zeta_m)}\int_0^a|\Psi(r)|^2J_0(\zeta_m
r/a)r\mathrm{d}r
\label{eq:pm}
\end{equation}

With knowledge of the displacement vectors we can readily calculate
the stresses and strains in the substrate via the standard relations
\cite{Landau86}.

The stresses and strains throughout the coating are a linear
combination of those in the substrate, (A4) of Harry \emph{et al.}
\cite{Harry02} give the necessary detail. We calculate the Lam\'{e}
parameters in the coating, taking account of the high and low index
materials, using average values for $Y$ and $\nu$. These averages
are calculated using the volumetric averaging operator introduced by
Fejer \cite{Fejer04}. We are now in a position to calculate the
magnitudes of the thermal noise in its various forms.

\subsection{Substrate Brownian Thermal Noise}
Brownian noise \cite{Bondu98} in solids is the thermally excited
motion associated with its intrinsic internal damping (i.e., damping
not associated with thermoelasticity). The key to reducing this
internal damping is to choose a substrate material having a small loss
angle $\Phi(\omega)$ or equivalently a high mechanical quality factor.

With estimates of $\Phi$ and having calculated the relevant stresses
and strains above we adopt the approach expounded in section V of Liu
and Thorne \cite{Liu00}. We only depart from their method in the
calculation of the expansion coefficients $p_m$ where we use the
values given by (\ref{eq:pm}).

\subsection{Substrate Thermoelastic Thermal Noise}

Thermoelastic noise is present in materials with a non-zero thermal expansion coefficient
undergoing thermal fluctuations \cite{Braginsky03}, and can dominate the thermal noise for certain
substrate materials.

To calculate the impact of this thermoelastic noise we again turn to
the arguments set forth by Liu \& Thorne \cite{Liu00}. The frequency at which we apply the fictitious
pressure to the test mass is much more rapid than the characteristic timescale for heat flow in the substrate,
so we may assume a
quasistatic system in which the temperature evolves adiabatically. Then the computation of
$W_{\mathrm{diss}}$ reduces to evaluating
\begin{equation}
  W_{\mathrm{diss}}=2\pi K_{\mathrm{ths}}T\bigg(\frac{Y\alpha}{(1-2\sigma)C_{\mathrm{v}}}\bigg)^2\frac{1}{2}\int_0^h\int_0^a(\nabla\Omega)^2r\mathrm{d}r\mathrm{d}z\\
 \end{equation}
 where $\Omega=\nabla\cdot\vec{u}$ is the divergence of the
 displacements calculated in \ref{sec:stressstrain}.

\subsection{Coating Brownian Thermal Noise}
We evaluate the Brownian noise in the coating using a model
developed by Harry \emph{et al.} \cite{Harry02}. This model allows
for the anisotropic layered structure of the coating and assumes
that the losses occur inside the materials themselves rather than at
the interfaces between coating layers. As stated for Brownian noise
in the substrate, the loss angle is of critical importance. In our
calculations we assume equality between loss angles parallel and
perpendicular to the coated surface, i.e. $\Phi_{||}=\Phi_{\bot}$ in
Harry's (15).

\subsection{Coating Thermoelastic Thermal Noise}
As with Brownian noise, thermoelastic noise is present in both the
substrate and the coating. One can simplify the calculation of the
coating contribution by assuming that the multi-layer coating may be
well-approximated by a uniform layer having appropriately averaged
properties.

Because the coating is very thin, the characteristic time scale for heat flow is short, and
we can no longer take it that heat flow normal to the
coating is adiabatic, hence we must adopt a perturbative approach. For
a comprehensive description of the methods used please see Fejer \emph{et al.}
\cite{Fejer04}. We embark on our analysis from a one dimensional (that
dimension normal to the coated surface) thermal conductivity
equation \cite{Lifshitz00}, our goal is to find the thermal field
$\Upsilon(z,t)$:

\begin{equation}
\frac{\partial\Upsilon_j}{\partial
t}-\kappa_j\frac{\partial^2\Upsilon_j}{\partial
z^2}=-\frac{Y_j\alpha_jT}{(1-2\sigma_j)C_{\mathrm{v}j}}\frac{\partial}{\partial
t}\sum_{i=1}^3E_{0,ii,j}
\end{equation}

Here the $j$ subscript acts as a place holder for $\mathrm{s}$ or $\mathrm{c}$ meaning
substrate or coating, we must solve this equation in both regions.
$\kappa_j=K_{\mathrm{th}j}/C_{\mathrm{v}j}$ is the thermal diffusivity and $E_{0,ii,j}$ is
the zeroth order (i.e. that due to the applied oscillatory Levin force)
$i$-polarised compressional strain.
We seek solutions enforcing the following:
\begin{eqnarray}
\Upsilon_{\mathrm{c}}|_{z=d}&=\Upsilon_{\mathrm{s}}|_{z=d}\\
K_{\mathrm{thc}}\frac{\partial\Upsilon_{\mathrm{c}}}{\partial z}\bigg|_{z=d}&
=K_{\mathrm{ths}}\frac{\partial\Upsilon_{\mathrm{s}}}{\partial z}\bigg|_{z=d}\\
\frac{\partial\Upsilon_{\mathrm{c}}}{\partial z}\bigg|_{z=0}&=0\\
\frac{\partial\Upsilon_{\mathrm{s}}}{\partial z}\bigg|_{z=h}&=0
\label{eq:finitemass}
\end{eqnarray}
i.e. continuity of temperature at the coating-substrate boundary and
zero heat flux at the surfaces of the test mass. Our equations are the
same as (17) in Fejer \cite{Fejer04} except for (\ref{eq:finitemass}),
where we assume a mirror of finite thickness, rather than the infinite
half-plane Fejer studies.  The substrate is so much thicker than the
coating that this leads to no quantitative difference in the result.

The equations are now simply solved and, applying the boundary
conditions, we have expressions for $\Upsilon_{\mathrm{c}}$ and
$\Upsilon_{\mathrm{s}}$.  Using the standard expression for the rate
of thermoelastic deformation
\begin{equation}
  W_{\mathrm{diss}}=\bigg\langle\int_{\mathrm{test\,mass}}\frac{K_{\mathrm{th}}}{T}(\nabla\delta T)^2\mathrm{d}V\bigg\rangle
  \label{eq:STE}
\end{equation}
with $\Upsilon=\delta T$ we can compute $W_{\mathrm{diss}}$ as follows
\begin{equation}
  \fl W_{\mathrm{diss}}=\frac{1}{2}\bigg(\frac{K_{\mathrm{thc}}}{T}2\pi\int_0^d\int_0^a|\nabla\Upsilon_{\mathrm{c}}|^2r\mathrm{d}r\mathrm{d}z+\frac{K_{\mathrm{th}}}{T}2\pi\int_d^h\int_0^a|\nabla\Upsilon_{\mathrm{s}}|^2r\mathrm{d}r\mathrm{d}z\bigg)
\end{equation}

\section{Material parameters}
\label{sec:MaterialParameters}
As far as possible we use the expected AdvLIGO values.

\begin{table}[htbp!]
\caption{The material parameters used in our calculations.}
\lineup
\begin{indented}
\item[]

\begin{tabular}{@{}llll}
\br
                        & Parameter                & Symbol       & Value\\
\mr
Substrate                       & Radius                                        & $r$                   & 0.17 m.\\
                                & Thickness                                     & $h$                   & 0.2 m.\\
                                & Density                                       & $\rho$                & 2.2$\times10^3$ kgm$^{-3}$\\
                                & Poisson ratio                                 & $\sigma$              & 0.17\\
                                & Young's modulus                               & $Y$                   & 7.2$\times10^{10}$ Nm$^{-2}$\\
                                & Loss angle                                    & $\Phi$                & $5\times10^{-9}$\\
                                & Linear thermal expansion coeff.               & $\alpha_{\mathrm{TH}}$         & 5.1$\times10^{-7}$ K$^{-1}$\\
                                & Thermal conductivity                          & $K_{\mathrm{th}}$                 & 1.38 Wm$^{-1}$K$^{-1}$\\
                                & Specific heat at const. volume                & $C_{\mathrm{v}}$                  & 1.64 $\times10^6$ JK$^{-1}$m$^{-3}$ \\
\mr
High $n$                        & Refractive index                              & $n_{\mathrm{ch}}$              & 2.03\\
material                        & Density                                       & $\rho_{\mathrm{ch}}$           & 6.85$\times10^3$ kgm$^{-3}$\\
                                & Poisson ratio                                 & $\sigma_{\mathrm{ch}}$         & 0.23\\
                                & Young's modulus                               & $Y_{\mathrm{ch}}$              & 1.4$\times10^{11}$ Nm$^{-2}$\\
                                & Loss angle                                    & $\Phi_{\mathrm{ch}}$           & $3.8\times10^{-4}$\\
                                & Linear thermal expansion coeff.               & $\alpha_{\mathrm{ch}}$         & 3.6$\times10^{-6}$ K$^{-1}$\\
                                & Thermal conductivity                          & $K_{\mathrm{thch}}$            & 33 Wm$^{-1}$K$^{-1}$\\
                                & Specific heat at const. volume                & $C_{\mathrm{vch}}$             & 2.1 $\times10^6$ JK$^{-1}$m$^{-3}$ \\
                                & Layer thickness ($\lambda_{LASER}/4n_{\mathrm{ch}}$)   & $d_{\mathrm{ch}}$              & 1.31 $\times10^{-7}$ m\\
\mr
         Low $n$                & Refractive index                              & $n_{\mathrm{cl}}$              & 1.45\\
         material               & Density                                       & $\rho_{\mathrm{cl}}$           & 2.2$\times10^3$ kgm$^{-3}$\\
                                & Poisson ratio                                 & $\sigma_{\mathrm{cl}}$         & 0.17\\
                                & Young's modulus                               & $Y_{\mathrm{cl}}$              & 7.2$\times10^{10}$ Nm$^{-2}$\\
                                & Loss angle                                    & $\Phi_{\mathrm{cl}}$           & $1\times10^{-4}$\\
                                & Linear thermal expansion coeff.               & $\alpha_{\mathrm{cl}}$         & 5.1$\times10^{-7}$ K$^{-1}$\\
                                & Thermal conductivity                          & $K_{\mathrm{thcl}}$            & 1.38 Wm$^{-1}$K$^{-1}$\\
                                & Specific heat at const. volume                & $C_{\mathrm{vcl}}$             & 1.64 $\times10^6$ JK$^{-1}$m$^{-3}$ \\
                                & Layer thickness ($\lambda_{\mathrm{LASER}}/4n_{\mathrm{cl}}$)   & $d_{\mathrm{cl}}$              & 1.83 $\times10^{-7}$ m\\
\br
\end{tabular}
\end{indented}
\end{table}

\begin{table}[htbp!]
\caption{Averaged material parameters used in the evaluation of
thermal noise in the SiO$_2$/Ta$_2$O$_5$ coating of a test mass.}
\lineup
\begin{indented}
\item[]
\begin{tabular}{@{}llll}
\br
                                & Parameter         & Symbol       & Value\\
\mr
Average                         & Poisson ratio                                 & $\sigma_{\mathrm{cavg}}$       & 0.195\\
coating                         & Young's modulus                               & $Y_{\mathrm{cavg}}$            & 1.003$\times10^{11}$ Nm$^{-2}$\\
values                          & Loss angle                                    & $\Phi_{\mathrm{cavg}}$         & $2.167\times10^{-4}$\\
                                & Linear thermal expansion coeff.               & $\alpha_{\mathrm{cavg}}$       & 1.798$\times10^{-6}$ K$^{-1}$\\
                                & Thermal conductivity                          & $K_{\mathrm{thcavg}}$          & 2.297 Wm$^{-1}$K$^{-1}$\\
                                & Specific heat at const. volume                & $C_{\mathrm{vcavg}}$           & 1.832 $\times10^6$ JK$^{-1}$m$^{-3}$ \\
                                & Total coating thickness                       & $d$                   & 5.975$\times10^{-6}$ m.\\

\br
\end{tabular}
\end{indented}
\end{table}

\vspace{3cm}
\section*{References}
\bibliographystyle{iopart-num}
 \bibliography{MesaBeamBibliography}

\end{document}